\documentclass[11pt]{article}
\usepackage[margin=1in]{geometry}
\usepackage[usenames,dvipsnames,svgnames,table]{xcolor} 

\usepackage{longtable}
\usepackage{hhline}
\usepackage{float}
\usepackage{booktabs}

\usepackage{setspace}
\usepackage{xspace}
\usepackage{enumitem} 
\usepackage{rotfloat} 
\usepackage{graphicx}
\usepackage{amssymb, amsmath, amsthm}
\usepackage{verbatim}
\usepackage{natbib}
\usepackage{authblk}
\usepackage{kotex}
\usepackage{multirow}
\usepackage{subcaption} 
\usepackage{array}
\usepackage{hyperref}
\newcolumntype{L}{>{\centering\arraybackslash}m{0.1\linewidth}}

\newtheorem{theorem}{Theorem}
\newtheorem{proposition}[theorem]{Proposition}

\newtheorem{corollary}[theorem]{Corollary}

\newcommand{\R}{\mathbb R}
\newcommand{\E}{\mathbb E}
\newcommand{\Pbb}{\mathbb P}
\newcommand{\Var}{\operatorname{Var}}

\newcommand{\tr}{\operatorname{tr}}

\newcommand{\argmin}{\operatorname*{arg\,min}}

\newcommand{\dto}{\rightsquigarrow}
\newcommand{\pto}{\stackrel{p}{\to}}

\newcommand{\Normal}{\mathcal N}
\newcommand{\M}{\mathcal M}
\newcommand{\U}{\mathcal U}
\newcommand{\V}{\mathcal V}
\newcommand{\Gr}{\operatorname{Gr}}

\newcommand{\Span}{\operatorname{span}}

\newcommand{\vecc}{\operatorname{vec}}

\newcommand{\eps}{\varepsilon}
\newcommand{\calL}{\mathcal L}

\begin{document}
	
\title{Scale-Calibrated Median-of-Means for Robust Distributed Principal Component Analysis}
\author[1,2]{Kisung You}
\affil[1]{Department of Mathematics, Baruch College}
\affil[2]{Department of Mathematics, The Graduate Center, City University of New York}
\date{}

\maketitle
\begin{abstract}
Distributed principal component analysis (PCA) produces node-level estimates of both
a mean vector and a principal subspace.  Robustly aggregating these heterogeneous
objects requires a relative scale between mean error and subspace error.  We
study a scale-calibrated median-of-means estimator for this problem using the
product geometry of Euclidean space and the Grassmann manifold.  A node-level
PCA expansion shows that the mean component has the usual linear influence,
whereas the subspace component is an eigengap-weighted covariance perturbation.
We prove a local reduction showing that the proposed product-manifold
median-of-means estimator is asymptotically equivalent to a scaled spatial
median of node influence errors.  This yields fixed-node non-Gaussian limits,
growing-node Gaussian limits with finite-block bias, and an explicit
scale-dependent covariance formula.  We propose robust block-scale and
inference-optimal calibration rules, establish high-probability
median-of-means bounds, characterize factorwise bad-node influence, and prove
node-bootstrap validity.  Simulations and large-scale single-cell RNA-seq data
show that scale calibration adapts to eigengap-driven subspace uncertainty and
provides a robust distributed PCA summary.
\end{abstract}

\section{Introduction}\label{sec:introduction}

Principal component analysis (PCA) is a basic tool for summarizing
high-dimensional data by a low-dimensional mean--subspace representation.  Since
the classical formulations of \citet{pearson_1901_LIIILinesPlanes} and
\citet{hotelling_1933_AnalysisComplexStatistical}, PCA has been used both as a
dimension-reduction method and as an inferential summary of dominant variation.
Its fixed-dimensional asymptotics and eigenspace perturbation behavior are well
understood
\citep{anderson_1963_AsymptoticTheoryPrincipal,
davis_1970_RotationEigenvectorsPerturbation,
stewart_1990_MatrixPerturbationTheory,
kato_1995_PerturbationTheoryLinear}.  These properties make PCA a natural
target for distributed computation: each node can compute a local PCA summary,
and only the local summaries need to be transmitted.

A local PCA summary, however, is not a single Euclidean object.  It consists of
a mean vector and a principal subspace.  The mean lies in \(\R^p\), whereas the
leading \(r\)-dimensional subspace lies on the Grassmann manifold
\(\Gr(r,p)\) \citep{edelman_1998_GeometryAlgorithmsOrthogonality}.  We use the
Grassmann formulation because PCA identifies a subspace, not a particular
signed, ordered, or oriented orthonormal basis.  A Stiefel representation would
introduce artificial orientation conventions and becomes unstable when
eigenvalues inside the leading block are tied or nearly tied.

The statistical stability of the two components can be very different.  A node
mean is governed by first-order variation of \(X\), whereas a node eigenspace is
governed by eigenprojector perturbation and is strongly affected by eigengaps.
When the leading eigenspace is weakly separated, the subspace component can be
much noisier than the mean component.  This imbalance is further amplified under
heavy tails or corrupted nodes.

Our goal is to robustly aggregate node-level PCA estimates
\[
    \widehat\theta_k=(\widehat\mu_k,\widehat\U_k)
    \in \R^p\times\Gr(r,p).
\]
Median-of-means aggregation is a natural way to combine robustness and
scalability
\citep{nemirovskij_1983_ProblemComplexityMethod,
jerrum_1986_RandomGenerationCombinatorial,
alon_1999_SpaceComplexityApproximatinga,
minsker_2015_GeometricMedianRobust,
lugosi_2019_SubGaussianEstimatorsMean}.  In the present problem, however, the
median must respect the product geometry of a Euclidean mean and a Grassmann
subspace.  This leads to the scale-calibrated product MoM estimator studied
below.

Once PCA is viewed this way, a scale issue becomes unavoidable.  Let \(B_1,\ldots,B_K\) be nodes of size \(b\), with \(n=Kb\), and let \(\widehat\theta_k=(\widehat\mu_k,\widehat\U_k)\) be the local PCA estimate.  We study the scaled geometric MoM estimator
\begin{equation}\label{eq:intro-estimator}
    \widetilde\theta_{n,\alpha}
    =\argmin_{(\mu,\U)\in\R^p\times\Gr(r,p)}
    {1\over K}\sum_{k=1}^K
    \left\{
    \alpha\|\mu-\widehat\mu_k\|^2
    +(2-\alpha)d_{\Gr}(\U,\widehat\U_k)^2
    \right\}^{1/2},
\end{equation}
where \(d_{\Gr}\) is the canonical Grassmann distance and \(\alpha\in(0,2)\).  The scale \(\alpha\) balances Euclidean mean error against subspace error, which determines how much a node must disagree in its mean or eigenspace before it is treated as unreliable by the final aggregation.

This paper studies how \(\alpha\) should be chosen and how it changes inference
in robust distributed PCA.  The question is related to, but distinct from,
ordinary scale selection for product-manifold medians.  For ordinary product
medians, changing factor weights can change the population center because the
objective couples factors through a common radial distance; fixed-scale
existence, robustness, limit theory, bootstrap validity, and algorithms are
available from \citet{park_2026_GeometricMediansProduct}.  Here, the aggregated
objects are node estimators, so scale acts on node-level errors.  In centered
MoM regimes, the target may remain \((\mu_0,\U_0)\), but \(\alpha\) changes
covariance, finite-block bias, concentration, confidence regions, and bad-node
influence.  Scale calibration is therefore an inferential design problem: small
eigengaps call for downweighting noisy Grassmann errors, whereas unstable mean
estimation calls for downweighting Euclidean errors. Our contributions are as follows.
\begin{enumerate}
\item We formulate robust distributed PCA as MoM estimation on
\(\R^p\times\Gr(r,p)\) and derive the joint influence expansion of
\((\widehat\mu_k,\widehat{\mathcal U}_k)\).  The mean influence is \(X-\mu_0\) and 
the Grassmann influence is an eigengap-weighted covariance perturbation.

\item We prove that the product-manifold MoM estimator
is asymptotically equivalent to a scaled spatial median of node influence
errors.  This yields both fixed-\(K\) non-Gaussian and growing-\(K\)
Gaussian limits with finite-block bias.

\item We derive the scale-dependent covariance of the centered aggregate.  If
\(H_\alpha\) is the block scale matrix and \(W\) is the limiting node influence
error, then
\begin{equation}\label{eq:intro-covariance}
    V_\alpha
    =
    H_\alpha^{-1/2}
    A_\alpha^{-1}
    S_\alpha
    A_\alpha^{-T}
    H_\alpha^{-1/2},
\end{equation}
where \(A_\alpha\) and \(S_\alpha\) are the derivative and score covariance of
the spatial median of \(H_\alpha^{1/2}W\).  This leads to robust block-scale and
inference-optimal calibration rules.

\item We establish high-probability MoM bounds, factorwise mean--subspace
deviation tradeoffs, bad-node robustness, scale-dependent influence constants
\(\alpha^{-1/2}\) and \((2-\alpha)^{-1/2}\), and node-bootstrap validity.
\end{enumerate}

The rest of the paper is organized as follows.  Section~\ref{sec:framework}
introduces the mean--subspace PCA parameter space
\(\R^p\times\Gr(r,p)\), defines the scaled product metric, and states the
standing assumptions.  Section~\ref{sec:local} derives the node-level PCA
influence expansion and proves the local reduction from product-manifold MoM
aggregation to a scaled spatial median of node errors.  Section~\ref{sec:limits}
develops the two asymptotic regimes: a fixed-\(K\) non-Gaussian limit and a
growing-\(K\) Gaussian limit with finite-block bias.  Section~\ref{sec:calibration}
studies scale-dependent inference, including the covariance formula, robust
block-scale calibration, inference-optimal scale selection, and the effect of
estimating \(\alpha\).  Section~\ref{sec:robustness} gives finite-sample
concentration, bad-node robustness, factorwise influence bounds, and node
bootstrap validity.  Section~\ref{sec-experiment} presents simulations and a
large-scale single-cell RNA-seq analysis.  Section~\ref{sec:conclusion}
concludes with implications and open directions.  Additional technical results,
all proofs, and supplementary numerical experiments are provided in the
\nameref{app:main}.


\section{Robust distributed PCA on \texorpdfstring{$\R^p\times\Gr(r,p)$}{R^p x Gr(r,p)}}\label{sec:framework}

We begin with the statistical object that motivates the paper. We note that while we study PCA, the later local MoM theory applies to more general product-valued estimators once a node-level asymptotic expansion is available.

\subsection{PCA target and Grassmann geometry}

Let \(X_1,\ldots,X_n\in\R^p\) be independent copies of a random vector \(X\).  Write
\begin{equation*}
    \mu_0=\E [X],
    \qquad
    \Sigma=\E[(X-\mu_0)(X-\mu_0)^T].
\end{equation*}
Denote  the eigendecomposition of \(\Sigma\) by
\[
    \Sigma=\sum_{j=1}^p \lambda_j u_j u_j^T,
    \qquad
    \lambda_1\ge\cdots\ge\lambda_p,
\]
with an eigengap
\begin{equation}\label{eq:eigengap}
    \lambda_r>\lambda_{r+1}.
\end{equation}
The leading \(r\)-dimensional principal subspace is
\[
    \U_0=\Span(u_1,\ldots,u_r)\in\Gr(r,p).
\]
The parameter of interest is
\[
    \theta_0=(\mu_0,\U_0)\in\M_p:=\R^p\times\Gr(r,p).
\]
The Grassmann distance is defined by principal angles.  If \(\vartheta_1,\ldots,\vartheta_r\in[0,\pi/2]\) are the principal angles between \(\U\) and \(\V\), then
\begin{equation}\label{eq:grassmann-distance}
    d_{\Gr}(\U,\V)^2=\sum_{j=1}^r \vartheta_j^2.
\end{equation}
This distance is invariant under the choice of orthonormal bases.  After choosing
an eigenbasis \(U_r=(u_1,\ldots,u_r)\) and an orthogonal complement \(U_\perp=(u_{r+1},\ldots,u_p)\), the tangent space \(T_{\U_0}\Gr(r,p)\) is identified with matrices in \(\R^{(p-r)\times r}\), equipped with the Frobenius norm.

For \(\alpha\in(0,2)\), define the scaled product distance
\begin{equation}\label{eq:scaled-distance}
    d_\alpha\{(\mu,\U),(\mu',\V)\}^2
    =\alpha\|\mu-\mu'\|^2+(2-\alpha)d_{\Gr}(\U,\V)^2.
\end{equation}
We will often restrict \(\alpha\) to a compact interval
\begin{equation}\label{eq:I-eps}
    I_\eps=[\eps,2-\eps],\qquad 0<\eps<1.
\end{equation}
This is not merely technical.  As later robustness bounds show, the factorwise influence constants diverge as \(\alpha\downarrow0\) or \(\alpha\uparrow2\).

\subsection{Local estimators and median aggregation}

For simplicity, we assume that \(\{1,\ldots,n\}\) can be partitioned into \(K\) disjoint nodes \(B_1,\ldots,B_K\) of equal size \(b\), with \(n=Kb\).  On node \(B_k\), define
\begin{equation}\label{eq:shard-mean}
    \widehat\mu_k={1\over b}\sum_{i\in B_k}X_i,
\end{equation}

a sample covariance
\begin{equation}\label{eq:shard-cov}
    \widehat\Sigma_k={1\over b}\sum_{i\in B_k}(X_i-\widehat\mu_k)(X_i-\widehat\mu_k)^T,
\end{equation}

and let \(\widehat\U_k\in\Gr(r,p)\) be the leading \(r\)-dimensional eigenspace of \(\widehat\Sigma_k\).  The local estimator is $\widehat\theta_k=(\widehat\mu_k,\widehat\U_k)$, and the scale-\(\alpha\) geometric MoM estimator is
\begin{equation}\label{eq:main-estimator}
    \widetilde\theta_{n,\alpha}
    =\argmin_{\theta\in\M_p} Q_{K,\alpha}(\theta),
    \qquad
    Q_{K,\alpha}(\theta)= {1\over K}\sum_{k=1}^K d_\alpha(\theta,\widehat\theta_k).
\end{equation}
For fixed \(\alpha\), this is a geometric median of points on a product manifold.  The general product-manifold theory therefore supplies the fixed-scale existence, local uniqueness, robustness, central limit, bootstrap, and computational foundations \citep{park_2026_GeometricMediansProduct}.  This paper does not revisit those fixed-scale product-median facts.  Instead, it studies how the scale \(\alpha\) interacts with the node-level PCA error distribution.

\subsection{Tangent notation and assumptions}\label{subsec:assumptions}

Let $d=p+r(p-r)$ be the dimension of the product tangent space.  In tangent coordinates at \(\theta_0=(\mu_0,\U_0)\), define
\begin{equation}\label{eq:H-alpha}
    H_\alpha=\begin{pmatrix}
        \alpha I_p & 0\\
        0 & (2-\alpha)I_{r(p-r)}
    \end{pmatrix}.
\end{equation}
For \(w=(w_\mu,w_\U)\in\R^p\oplus T_{\U_0}\Gr(r,p)\), write
\[
    \|w\|_{H_\alpha}^2=w^T H_\alpha w
    =\alpha\|w_\mu\|^2+(2-\alpha)\|w_\U\|^2.
\]

We use the following assumptions throughout the theoretical development.  The first two conditions ensure that the node PCA estimators are well-defined in a common local coordinate system.  The third condition is the usual regularity condition for spatial-median asymptotics applied to the node-level errors.  The final condition is only needed for refined growing-\(K\) inference with finite-block bias.

\begin{enumerate}[label=(A\arabic*), ref=(A\arabic*), leftmargin=2.7em]
\item \label{ass:pca} \textit{Moments and eigengap.}
The observations satisfy
\[    \E\|X-\mu_0\|^{4+\delta}<\infty
\]
for some \(\delta>0\), and the population covariance matrix satisfies the
eigengap condition \eqref{eq:eigengap}.  The eigenvalues \(\lambda_j\) may have
multiplicities within the leading block and within the trailing block. Only the
separation between the leading \(r\)-dimensional eigenspace and its orthogonal
complement is required.
\item \label{ass:localization} \textit{Node localization.} There is a normal neighborhood \(\mathcal N\) of \(\U_0\) in \(\Gr(r,p)\) such that \(\max_{1\le k\le K}\mathbf 1\{\widehat\U_k\notin\mathcal N\}\to0\) in probability.  For fixed \(K\), this follows from \(\widehat\U_k\to\U_0\) in probability.  For growing \(K\), it is enough to assume
\[
    K\,\Pbb(\widehat\U_1\notin\mathcal N)\to0.
\]
The same neighborhood is used for all \(\alpha\in I_\eps\).
\item \label{ass:spatial-regularity} \textit{Spatial median regularity.} Let
\begin{equation}\label{eq:Wkb-def}
    W_{k,b}=\sqrt b\begin{pmatrix}
        \widehat\mu_k-\mu_0\\
        \log_{\U_0}(\widehat\U_k)
    \end{pmatrix}
\end{equation}
be the node error in product tangent coordinates, and let \(W_b\) denote a generic copy.  For each \(\alpha\in I_\eps\), define
\begin{equation}\label{eq:s-alpha-b-def}
    s_{\alpha,b}=\argmin_{s\in\R^d}\E\|s-W_b\|_{H_\alpha}.
\end{equation}
The minimizer is unique.  With
\[
    Y_{\alpha,b}=H_\alpha^{1/2}(W_b-s_{\alpha,b}),
    \quad
    R_{\alpha,b}=\|Y_{\alpha,b}\|,
    \quad
    U_{\alpha,b}=Y_{\alpha,b}/R_{\alpha,b},
\]
we assume \(\Pbb(R_{\alpha,b}=0)=0\), \(\E R_{\alpha,b}^{-1}<\infty\), and
\begin{equation}\label{eq:A-alpha-b-def}
    A_{\alpha,b}=\E\left[{I-U_{\alpha,b}U_{\alpha,b}^T\over R_{\alpha,b}}\right]
\end{equation}
 is nonsingular uniformly over \(\alpha\in I_\eps\).
\item \label{ass:bias} \textit{Finite-block bias expansion.} For refined growing-$K$  inference, the finite-block median satisfies
\begin{equation}\label{eq:bias-expansion}
    s_{\alpha,b}=b^{-1/2}a_\alpha+o(b^{-1/2})
\end{equation}
uniformly over \(\alpha\in I_\eps\), for a continuous function \(\alpha\mapsto a_\alpha\).
\end{enumerate}

The finite-block bias \(s_{\alpha,b}\) is the population spatial median of the
finite-sample node error distribution.  It vanishes under central symmetry, but
otherwise can be of order \(b^{-1/2}\).  Hence it may survive root-\(n\)
scaling unless \(K/b\to0\).

\section{Local PCA asymptotics and MoM reductions}\label{sec:local}

In this section, we convert robust distributed PCA into a problem about spatial medians in a Euclidean tangent space.  This reduction allows the later covariance and robustness results to be expressed in explicit forms.  It proceeds in two steps.  First, we linearize the node PCA estimator.  Second, we show that the geometric median of node estimators is locally the spatial median of the linearized errors.

\subsection{Influence expansion for node PCA}

Let
\[
    E_X=(X-\mu_0)(X-\mu_0)^T-\Sigma.
\]
Choose the eigenbasis \(u_1,\ldots,u_p\) used in the decomposition of \(\Sigma\).  For \(a=1,
\ldots,p-r\) and \(j=1,\ldots,r\), define
\begin{equation}\label{eq:B-influence}
    B(X)_{aj}
    ={u_{r+a}^T E_X u_j\over \lambda_j-\lambda_{r+a}}.
\end{equation}
This matrix is the first-order tangent representation of the leading subspace perturbation.  The denominators are eigengaps between the leading eigenspace and its complement.  Thus the formula already reveals the statistical difficulty of PCA such that subspace uncertainty is amplified when the leading and trailing eigenvalues are close.

\begin{theorem}[Node PCA influence expansion]\label{thm:pca-influence}
Suppose Assumption \ref{ass:pca} holds.  For a node \(B_k\), define \(W_{k,b}\) as in \eqref{eq:Wkb-def}.  Then
\begin{equation}\label{eq:pca-influence}
    W_{k,b}
    ={1\over\sqrt b}\sum_{i\in B_k}\zeta(X_i)+r_{k,b},
    \qquad
    r_{k,b}\pto0,
\end{equation}
where
\begin{equation}\label{eq:zeta-def}
    \zeta(X)=\begin{pmatrix}
        X-\mu_0\\
        \vecc\{B(X)\}
    \end{pmatrix}\in\R^{p+r(p-r)}.
\end{equation}
Consequently,
\begin{equation}\label{eq:W-clt}
    W_{k,b}\dto W\sim\Normal(0,\Gamma),
    \qquad
    \Gamma=\Var\{\zeta(X)\}.
\end{equation}
If \(K=K_n\to\infty\), the expansion is uniform over nodes whenever
\begin{equation}\label{eq:uniform-remainder-condition}
    \sqrt K\max_{1\le k\le K}\|r_{k,b}\|\pto0.
\end{equation}
\end{theorem}

The theorem is a first-order spectral perturbation expansion written in product tangent coordinates.  The mean part is linear, while the Grassmann part is an eigengap-weighted covariance perturbation.  Thus the covariance \(\Gamma\) contains mean variability, subspace variability, and their cross-covariance, which are precisely the quantities later affected by scale.

\subsection{Local reduction to a scaled spatial median}

The next result explains why the product-manifold aggregation can be studied through the spatial median of \(W_{k,b}\).  Define
\begin{equation}\label{eq:tangent-median}
    \widehat s_{K,b,\alpha}
    =\argmin_{s\in\R^d}{1\over K}\sum_{k=1}^K\|s-W_{k,b}\|_{H_\alpha}.
\end{equation}
This is the scaled spatial median of the node PCA errors in tangent coordinates, which corresponds to the first-order representation of the actual geometric MoM  estimator.

\begin{theorem}[Local reduction]\label{thm:local-reduction}
Suppose Assumptions \ref{ass:pca}, \ref{ass:localization}, and \ref{ass:spatial-regularity} hold.  If \(K\) is fixed, then uniformly over \(\alpha\in I_\eps\),
\begin{equation}\label{eq:local-reduction-fixed}
    \sqrt b\log_{\theta_0}\widetilde\theta_{n,\alpha}
    =\widehat s_{K,b,\alpha}+o_p(1).
\end{equation}
If \(K\to\infty\), the stronger relation
\begin{equation}\label{eq:local-reduction-growing}
    \sqrt K\sup_{\alpha\in I_\eps}
    \left\|
    \sqrt b\log_{\theta_0}\widetilde\theta_{n,\alpha}
    -\widehat s_{K,b,\alpha}
    \right\|
    \pto0
\end{equation}
 holds provided the local Riemannian objective approximation error is \(o_p(K^{-1/2})\) uniformly over \(\alpha\in I_\eps\).  A sufficient condition is \(\sqrt K/b\to0\) together with a uniformly bounded third moment for the localized node errors.
\end{theorem}

The theorem reduces the nonlinear product-manifold aggregation problem to an ordinary spatial median problem in \(\R^d\).  All scale-dependent covariance and robustness calculations below are consequences of this tangent-space representation.

\section{Limit theory for scale-calibrated MoM PCA}\label{sec:limits}

The local reduction separates two sources of asymptotics: the within-node limit that creates the error distribution \(W\), and the across-node median that aggregates \(K\) errors.  These two sources lead to two regimes.  If \(K\) is fixed, the limit is a finite geometric median of random vectors and is generally non-Gaussian.  If \(K\to\infty\), the spatial median itself has a central limit theorem, but the finite-block population median \(s_{\alpha,b}\) can contribute a bias term.

\subsection{Fixed number of nodes}

\begin{theorem}[Fixed-\(K\) limit]\label{thm:fixed-K}
Suppose \(K\) is fixed and the conditions of Theorems \ref{thm:pca-influence} and \ref{thm:local-reduction} hold.  Let \(W_1,\ldots,W_K\) be independent copies of \(W\sim\Normal(0,\Gamma)\).  If
\begin{equation}\label{eq:T-alpha-K}
    T_{\alpha,K}=\argmin_{s\in\R^d}{1\over K}\sum_{k=1}^K\|s-W_k\|_{H_\alpha}
\end{equation}
 is unique almost surely, then
\begin{equation}\label{eq:fixed-K-limit}
    \sqrt b\log_{\theta_0}\widetilde\theta_{n,\alpha}
    \dto T_{\alpha,K}.
\end{equation}
Equivalently,
\begin{equation}\label{eq:fixed-K-root-n}
    \sqrt n\log_{\theta_0}\widetilde\theta_{n,\alpha}
    \dto \sqrt K\,T_{\alpha,K}.
\end{equation}
\end{theorem}

This regime is relevant when the number of machines is fixed.  The estimator is root-\(n\), but the limit is generally non-Gaussian because it is the spatial median of finitely many Gaussian node errors.

\subsection{Growing number of nodes}

When \(K\to\infty\), the spatial median of the node errors has its own central limit theorem.  The centering, however, is the finite-block population median \(s_{\alpha,b}\), not necessarily zero.  This distinction is essential as MoM has two asymptotic levels, and the finite-block asymmetry of the node estimator can survive root-\(n\) scaling if \(K\) grows too quickly relative to \(b\).

Define
\begin{equation}\label{eq:S-alpha-b}
    S_{\alpha,b}=\Var(U_{\alpha,b}),
\end{equation}
where \(U_{\alpha,b}\) was defined in Assumption \ref{ass:spatial-regularity}.  The covariance associated with the spatial median of finite-block errors is
\begin{equation}\label{eq:V-alpha-b}
    V_{\alpha,b}
    =H_\alpha^{-1/2}A_{\alpha,b}^{-1}S_{\alpha,b}A_{\alpha,b}^{-T}H_\alpha^{-1/2}.
\end{equation}

\begin{theorem}[Growing-\(K\) CLT with finite-block bias]\label{thm:growing-K}
Suppose \(K\to\infty\), \(b\to\infty\), and \(n=Kb\).  Under Assumptions \ref{ass:pca}--\ref{ass:spatial-regularity} and the growing-\(K\) reduction condition in \eqref{eq:local-reduction-growing},
\begin{equation}\label{eq:spatial-clt-finite-b}
    \sqrt K\left(\widehat s_{K,b,\alpha}-s_{\alpha,b}\right)
    \dto \Normal(0,V_{\alpha,b}).
\end{equation}
Consequently,
\begin{equation}\label{eq:mom-bias-expansion}
    \sqrt n\log_{\theta_0}\widetilde\theta_{n,\alpha}
    =\sqrt K\,s_{\alpha,b}+Z_{\alpha,b}+o_p(1),
\end{equation}
where \(Z_{\alpha,b}\dto\Normal(0,V_\alpha)\) whenever \(V_{\alpha,b}\to V_\alpha\).  If \(\sqrt K\,s_{\alpha,b}\to\delta_\alpha\), then
\begin{equation}\label{eq:growing-K-limit}
    \sqrt n\log_{\theta_0}\widetilde\theta_{n,\alpha}
    \dto \Normal(\delta_\alpha,V_\alpha).
\end{equation}
In particular, if \(\sqrt K\,s_{\alpha,b}\to0\), then the limit is centered.
\end{theorem}

\begin{corollary}[Symmetry and the node growth condition]\label{cor:symmetry-Kb}
If the finite-block node error distribution is centrally symmetric, \(W_b\stackrel d=-W_b\), then \(s_{\alpha,b}=0\) for every \(\alpha\in I_\eps\), and the centered CLT holds.  More generally, if Assumption \ref{ass:bias} holds, then
\[
    \sqrt K\,s_{\alpha,b}=\sqrt{K/b}\,a_\alpha+o(\sqrt{K/b}).
\]
Thus, absent symmetry or cancellation of \(a_\alpha\), centered root-\(n\) inference requires
\begin{equation}\label{eq:K-over-b}
    K/b\to0.
\end{equation}
\end{corollary}

Corollary~\ref{cor:symmetry-Kb} gives a concrete design lesson.  Increasing \(K\) improves robustness to bad nodes and parallel scalability, but it reduces the node size \(b\).  If the node PCA error has finite-block skewness, the product median aggregation can have a root-\(n\) centering bias unless the nodes remain large enough.  This indicates the nature of \(K\), \(b\), and \(\alpha\) as inferential design choices rather than purely computational parameters.

\section{Scale-dependent inference and calibration}\label{sec:calibration}

We now turn from limit laws to scale selection.  The first result gives the limiting covariance as a function of \(\alpha\).  The remaining results use that covariance to justify block-noise calibration, robust empirical scale estimation, and inference-optimal tuning.

\subsection{The covariance induced by a scale}

Assume the limiting node error is \(W\sim\Normal(0,\Gamma)\).  For \(\alpha\in I_\eps\), define
\[
    Y_\alpha=H_\alpha^{1/2}W,
    \qquad
    R_\alpha=\|Y_\alpha\|,
    \qquad
    U_\alpha=Y_\alpha/R_\alpha.
\]
If the spatial median of \(Y_\alpha\) is zero, set
\begin{equation}\label{eq:A-alpha}
    A_\alpha=\E\left[{I-U_\alpha U_\alpha^T\over R_\alpha}\right],
    \qquad
    S_\alpha=\Var(U_\alpha).
\end{equation}

\begin{theorem}[Scale-dependent covariance]\label{thm:V-alpha}
Under the centered growing-\(K\) regime of Theorem \ref{thm:growing-K}, if \(W_b\dto W\sim\Normal(0,\Gamma)\), the spatial median of \(Y_\alpha=H_\alpha^{1/2}W\) is zero, and \(A_\alpha\) is nonsingular, then
\begin{equation}\label{eq:V-alpha}
    V_\alpha
    =H_\alpha^{-1/2}A_\alpha^{-1}S_\alpha A_\alpha^{-T}H_\alpha^{-1/2}.
\end{equation}
\end{theorem}

The formula shows exactly how \(\alpha\) enters inference.  The scale matrix first transforms the node error distribution from \(W\) to \(Y_\alpha=H_\alpha^{1/2}W\).  The geometric median is most efficient when this transformed distribution is close to spherical.  The final factor \(H_\alpha^{-1/2}\) converts the covariance back to the original mean-subspace tangent coordinates.

For PCA, the covariance \(\Gamma\) decomposes as
\begin{equation}\label{eq:Gamma-blocks}
    \Gamma=\begin{pmatrix}
        \Gamma_{\mu\mu} & \Gamma_{\mu\U}\\
        \Gamma_{\U\mu} & \Gamma_{\U\U}
    \end{pmatrix}.
\end{equation}
The block \(\Gamma_{\mu\mu}\) describes mean-estimation uncertainty; \(\Gamma_{\U\U}\) describes eigenspace-estimation uncertainty; and the cross-blocks describe dependence between the two.  Since \(\Gamma_{\U\U}\) contains eigengap denominators through \eqref{eq:B-influence}, weak eigengaps can strongly affect \(V_\alpha\).

\subsection{Whitening and block-noise calibration}

A simple but useful case occurs when the mean and subspace blocks are approximately isotropic and weakly correlated:
\[
    \Gamma\approx\begin{pmatrix}
        \sigma_\mu^2 I_p & 0\\
        0 & \sigma_\U^2 I_{r(p-r)}
    \end{pmatrix}.
\]
Then the transformed covariance \(H_\alpha^{1/2}\Gamma H_\alpha^{1/2}\) is spherical when
\[
    \alpha\sigma_\mu^2=(2-\alpha)\sigma_\U^2,
\]
which gives
\begin{equation}\label{eq:alpha-white}
    \alpha_{\rm white}=2{\sigma_\U^2\over \sigma_\mu^2+\sigma_\U^2}.
\end{equation}
More generally, define average block variances
\begin{equation}\label{eq:tau-blocks}
    \tau_\mu={1\over p}\tr(\Gamma_{\mu\mu}),
    \qquad
    \tau_\U={1\over r(p-r)}\tr(\Gamma_{\U\U}).
\end{equation}
The block-noise calibration rule is
\begin{equation}\label{eq:alpha-block}
    \alpha_{\rm block}=2{\tau_\U\over \tau_\mu+\tau_\U}.
\end{equation}
This rule calibrates the metric to the uncertainty of the node estimators, not to raw data dispersion.  In PCA terms, if the subspace estimator is noisy, then \(\tau_\U\) is large and \(\alpha_{\rm block}\) moves toward two, thereby downweighting the Grassmann term \(2-\alpha\).  If the mean estimator is noisy, \(\alpha_{\rm block}\) moves toward zero, downweighting the Euclidean term.

\begin{proposition}[Efficiency under perfect whitening]\label{prop:efficiency-whitening}
If \(Y_\alpha\sim\Normal(0,\sigma^2I_d)\), with \(d>1\), then
\begin{equation}\label{eq:cd-constant}
    A_\alpha^{-1}S_\alpha A_\alpha^{-T}=c_d\sigma^2I_d,
    \qquad
    c_d={2d\,\Gamma(d/2)^2\over (d-1)^2\Gamma((d-1)/2)^2}.
\end{equation}
Moreover, \(c_d\to1\) as \(d\to\infty\).
\end{proposition}

This proposition gives the efficiency intuition behind block-noise calibration.  A spatial median is generally less efficient than a mean under exact Gaussianity, but when the transformed node errors are spherical the loss is summarized by a scalar constant that approaches one in high dimension.  Calibration can therefore recover much of the efficiency while retaining median-of-means robustness to corrupted nodes.

\subsection{Robust empirical scale calibration}

The covariance \(\Gamma\) can be estimated from node errors, but covariance estimates may be unstable when some nodes are corrupted.  We therefore use a robust radial calibration.  For a preliminary estimator, e.g. \(\widetilde\theta_{\rm prelim}=\widetilde\theta_{n,1}\), define
\begin{equation}\label{eq:smu-hat}
    \widehat s_\mu=	\operatorname{median}_{1\le k\le K}
    \left\{\sqrt b\,\|\widehat\mu_k-\widetilde\mu_{\rm prelim}\|\right\},
\end{equation}
\begin{equation}\label{eq:sU-hat}
    \widehat s_\U=	\operatorname{median}_{1\le k\le K}
    \left\{\sqrt b\,d_{\Gr}(\widehat\U_k,\widetilde\U_{\rm prelim})\right\}.
\end{equation}
For per-tangent-dimension calibration, set
\begin{equation}\label{eq:tau-hats}
    \widehat\tau_\mu={\widehat s_\mu^2\over p},
    \qquad
    \widehat\tau_\U={\widehat s_\U^2\over r(p-r)},
\end{equation}
 and define
\begin{equation}\label{eq:alpha-rpca}
    \widehat\alpha_{\rm rPCA}
    =2{\widehat\tau_\U\over \widehat\tau_\mu+\widehat\tau_\U}.
\end{equation}

\begin{theorem}[Consistency of robust PCA scale calibration]\label{thm:scale-consistency}
Suppose \(\widehat s_\mu\pto s_\mu\) and \(\widehat s_\U\pto s_\U\), where \(s_\mu,s_\U\in(0,\infty)\).  Then
\begin{equation}\label{eq:alpha-rpca-limit}
    \widehat\alpha_{\rm rPCA}\pto
    \alpha_{\rm rPCA}
    =2{ s_\U^2/[r(p-r)]\over s_\mu^2/p+s_\U^2/[r(p-r)] }.
\end{equation}
If the limiting mean and subspace errors are block-isotropic, then \(\alpha_{\rm rPCA}\) agrees with the whitening scale up to the known dimension-dependent radial constants of the block norms.
\end{theorem}

\subsection{Inference-optimal scale and the effect of estimating scale}

Calibration by radial block scales is simple and robust.  A more directly inferential alternative is to choose \(\alpha\) by minimizing a covariance risk.  Let
\begin{equation}\label{eq:risk-alpha}
    R(\alpha)=\tr(W_0V_\alpha),
\end{equation}
for a user-chosen positive semidefinite weight matrix \(W_0\).  Other choices include \(R(\alpha)=\log\det V_\alpha\) or \(R(\alpha)=\lambda_{\max}(V_\alpha)\).  Let
\begin{equation}\label{eq:alpha-star}
    \alpha^*=\argmin_{\alpha\in I_\eps}R(\alpha).
\end{equation}
Estimate \(\alpha^*\) by replacing \(V_\alpha\) with a plug-in or bootstrap estimate \(\widehat V_\alpha\).

\begin{theorem}[Inference-optimal scale]\label{thm:optimal-scale}
Assume \(\sup_{\alpha\in I_\eps}\|\widehat V_\alpha-V_\alpha\|\pto0\) and \(R\) has a unique minimizer \(\alpha^*\) in \(I_\eps\).  Then
\begin{equation}\label{eq:alpha-star-consistency}
    \widehat\alpha^*=\argmin_{\alpha\in I_\eps}\widehat R(\alpha)
    \pto \alpha^*.
\end{equation}
If \(R\) is twice continuously differentiable at \(\alpha^*\), \(R''(\alpha^*)>0\), and \(\sqrt K\{\widehat R'(\alpha^*)-R'(\alpha^*)\}\dto N(0,\sigma_R^2)\), then
\begin{equation}\label{eq:alpha-star-clt}
    \sqrt K(\widehat\alpha^*-\alpha^*)\dto N\left(0,{\sigma_R^2\over R''(\alpha^*)^2}\right).
\end{equation}
\end{theorem}

A practical question is whether estimating \(\alpha\) changes the first-order distribution of the final estimator.  The answer depends on the finite-block median path \(s_{\alpha,b}\).  If this path is zero, scale estimation affects the limiting covariance through the selected limit \(\alpha_0\) but does not add a first-order correction.  If the finite-block path is nonzero, changes in \(\alpha\) can also move the centering.

\begin{theorem}[First-order effect of estimating scale]\label{thm:estimated-alpha}
Let \(\widehat\alpha\pto\alpha_0\in I_\eps\).  Suppose the median process \(\sqrt K(\widehat s_{K,b,\alpha}-s_{\alpha,b})\) is stochastically equicontinuous in \(\alpha\).  If \(s_{\alpha,b}=0\) for all \(\alpha\in I_\eps\), then
\begin{equation}\label{eq:estimated-alpha-centered}
    \sqrt n\log_{\theta_0}\widetilde\theta_{n,\widehat\alpha}
    \dto \Normal(0,V_{\alpha_0}).
\end{equation}
If \(s_{\alpha,b}\) is nonzero, then the expansion contains the additional centering term
\begin{equation}\label{eq:estimated-alpha-bias}
    \sqrt K\{s_{\widehat\alpha,b}-s_{\alpha_0,b}\}.
\end{equation}
In particular, if \(s_{\alpha,b}=b^{-1/2}a_\alpha+o(b^{-1/2})\) uniformly and \(\widehat\alpha-\alpha_0=O_p(K^{-1/2})\), then \eqref{eq:estimated-alpha-bias} is \(O_p(b^{-1/2})\) and is negligible as \(b\to\infty\).
\end{theorem}

\section{Concentration, robustness, and resampling}\label{sec:robustness}

The asymptotic results show how scale affects limiting inference.  Median-of-means, however, is valued because it offers finite-sample robustness to corrupted nodes.  This section gives deterministic and probabilistic guarantees that make the role of scale explicit.

\subsection{Good-node concentration}

We revisit the geometric core of MoM robustness, closely related to the majority-ball arguments used in robust mean estimation, that is deterministic and holds in any metric space. \citep{minsker_2015_GeometricMedianRobust, lugosi_2019_SubGaussianEstimatorsMean}.

\begin{theorem}[Deterministic good-node lemma]\label{thm:good-node}
Let \(z_1,
\ldots,z_K\in\M_p\), and let
\[
    \widetilde z_\alpha\in\argmin_{z\in\M_p}\sum_{k=1}^K d_\alpha(z,z_k).
\]
Suppose at least \((1/2+\gamma)K\) of the points satisfy \(d_\alpha(z_k,\theta_0)\le r\), where \(\gamma\in(0,1/2]\).  Then
\begin{equation}\label{eq:good-node-bound}
    d_\alpha(\widetilde z_\alpha,\theta_0)
    \le \left(1+{1\over 2\gamma}\right)r.
\end{equation}
\end{theorem}

Combining this lemma with a node-level second-moment bound gives the usual MoM exponential tail form, now with the scaled product metric.

\begin{theorem}[Scaled high-probability MoM bound]\label{thm:high-prob}
Assume that for every node
\begin{equation}\label{eq:node-second-moment}
    \E d_\alpha(\widehat\theta_k,\theta_0)^2
    \le {\alpha v_\mu+(2-\alpha)v_\U\over b}
\end{equation}
for constants \(v_\mu,v_\U>0\).  Then there exist universal constants \(c_1,c_2>0\) such that
\begin{equation}\label{eq:scaled-high-prob}
    \Pbb\left(
    d_\alpha(\widetilde\theta_{n,\alpha},\theta_0)
    > c_1\sqrt{\alpha v_\mu+(2-\alpha)v_\U\over b}
    \right)
    \le \exp(-c_2K).
\end{equation}
\end{theorem}

The theorem gives a bound in the scaled product metric, which leads to the following corollary that translates into the mean and the principal subspace.

\begin{corollary}[Factorwise deviation tradeoff]\label{cor:factorwise}
Under the conditions of Theorem \ref{thm:high-prob}, with probability at least \(1-\exp(-c_2K)\),
\begin{equation}\label{eq:mean-factor-bound}
    \|\widetilde\mu_{n,\alpha}-\mu_0\|
    \le {c_1\over\sqrt\alpha}
    \sqrt{\alpha v_\mu+(2-\alpha)v_\U\over b},
\end{equation}
 and
\begin{equation}\label{eq:subspace-factor-bound}
    d_{\Gr}(\widetilde\U_{n,\alpha},\U_0)
    \le {c_1\over\sqrt{2-\alpha}}
    \sqrt{\alpha v_\mu+(2-\alpha)v_\U\over b}.
\end{equation}
\end{corollary}

This corollary is the finite-sample analogue of the covariance tradeoff.  Moving \(\alpha\) toward two may reduce the scaled effect of subspace noise, but it also weakens the direct subspace guarantee because the factor \((2-\alpha)^{-1/2}\) grows.  Moving \(\alpha\) toward zero has the analogous effect on the mean.

\subsection{Bad nodes and factorwise influence}

The median aggregation tolerates arbitrary corrupted nodes as long as fewer than half the nodes are bad.  This breakdown threshold does not depend on \(\alpha\).  The direction and magnitude of factorwise influence do depend on \(\alpha\).  In tangent coordinates, define the node score
\begin{equation}\label{eq:psi-alpha}
    \psi_\alpha(w)={w\over \|w\|_{H_\alpha}},
    \qquad w=(w_\mu,w_\U)\ne0.
\end{equation}

\begin{theorem}[Bad-node robustness and factorwise influence]\label{thm:influence}
If fewer than half of the node estimators are arbitrary and the remaining nodes are contained in a bounded \(d_\alpha\)-ball around \(\theta_0\), then the scaled median-of-means estimator remains bounded.  Moreover,
\begin{equation}\label{eq:influence-mean}
    \sup_{w\ne0}\|\psi_{\alpha,\mu}(w)\|=\alpha^{-1/2},
\end{equation}
 and
\begin{equation}\label{eq:influence-subspace}
    \sup_{w\ne0}\|\psi_{\alpha,\U}(w)\|=(2-\alpha)^{-1/2}.
\end{equation}
Thus the breakdown fraction is scale-invariant, but the mean and subspace influence constants are scale-dependent.
\end{theorem}

This result is practically important for our proposal.  A node can be corrupted primarily in the mean while its subspace estimate remains plausible, or vice versa.  The scale determines how visible such corruption is to the radial denominator in the median score.

\subsection{Node bootstrap}

For inference, one may either estimate \(V_\alpha\) by plug-in formulas or resample the node estimators.  In this paper, we consider the node bootstrap, which is especially natural in distributed settings because it does not require resampling raw observations.

Let \(\widehat\theta_1^*,\ldots,\widehat\theta_K^*\) be sampled with replacement from \(\widehat\theta_1,
\ldots,\widehat\theta_K\), and define \(\widetilde\theta_{n,\alpha}^*\) by applying the same scaled median aggregation to the bootstrap node estimates.

\begin{theorem}[Node bootstrap validity]\label{thm:bootstrap}
Under the centered growing-\(K\) conditions of Theorem \ref{thm:growing-K}, conditionally on the observed nodes,
\begin{equation}\label{eq:bootstrap-validity}
    \sqrt n\log_{\widetilde\theta_{n,\alpha}}
    \widetilde\theta_{n,\alpha}^*
    \dto \Normal(0,V_\alpha)
\end{equation}
 in probability.  If \(\alpha\) is estimated and the negligibility condition of Theorem \ref{thm:estimated-alpha} is not invoked, then the bootstrap procedure should recompute \(\widehat\alpha^*\) inside each bootstrap replicate.
\end{theorem}

\section{Experiments}\label{sec-experiment}
\subsection{Eigengap-driven subspace uncertainty}
\label{subsec:sim-eigengap}

The first simulation examines whether the proposed scale calibration responds to the main statistical difficulty in PCA: instability of the leading eigenspace when the eigengap is small.  We generated data from a spiked covariance model in \(\R^p\) with \(p=200\) and target rank \(r=5\).  The sample size was \(n=40000\), split into \(K=80\) equal-sized nodes, and results were averaged over 60 Monte Carlo replications.  The eigengap $\Delta=\lambda_r-\lambda_{r+1}$ was varied over
$\Delta\in\{0.1,0.25,0.5,1,2,4\}.$ For each node, we computed the node mean and leading \(r\)-dimensional principal subspace.  We compared full-sample PCA, random subset PCA using one node-sized subsample, projector-averaged distributed PCA, fixed-scale MoM PCA with \(\alpha=1\), and the proposed robust scale-calibrated MoM PCA.

This experiment is designed to isolate the role of the eigengap.  The influence expansion shows that the Grassmann component contains eigengap-weighted covariance perturbations.  Thus, as \(\Delta\) decreases, the node subspace estimates become more variable even when the node mean estimates remain stable.  A useful scale calibration rule should detect this imbalance from node-level errors and downweight the Grassmann component when the subspace is weakly separated.

\begin{figure}[tb]
    \centering
    \includegraphics[width=\textwidth]{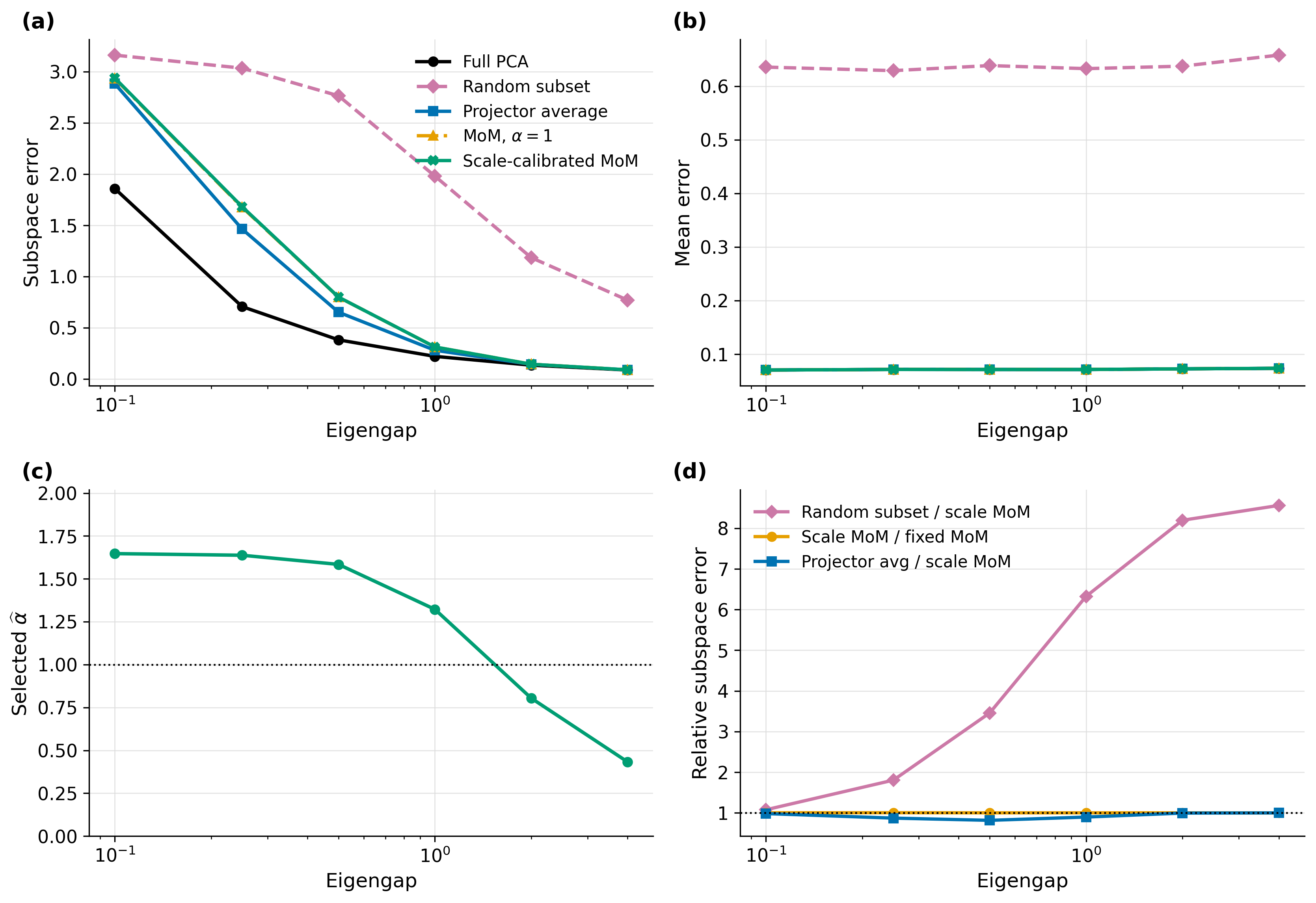}
    \caption{
Spiked PCA simulation with \(p=200\), \(r=5\), \(n=40000\), and \(K=80\) nodes. 
As the eigengap increases, subspace estimation becomes easier and the calibrated scale adjusts to the changing mean--subspace uncertainty. 
Panels show (a) Grassmann subspace error, (b) Euclidean mean error, (c) selected scale \(\widehat\alpha_{\rm rPCA}\), and (d) subspace error relative to scale-calibrated MoM.
}
    \label{fig:sim1-eigengap}
\end{figure}

Figure~\ref{fig:sim1-eigengap} summarizes the estimation performance.  The random subset baseline is unstable across all eigengaps, especially in the subspace coordinate.  At \(\Delta=1\), for example, the average subspace error of the random subset estimator is about \(1.98\), compared with \(0.31\) for the MoM estimators and \(0.28\) for projector averaging.  Full-sample PCA is the clean-data reference and is, as expected, the most accurate estimator.  Projector averaging is also efficient in this clean Gaussian setting.  The MoM estimators pay a modest clean-efficiency cost relative to projector averaging, but remain far more stable than using a single random subset.  This is the expected price of robust aggregation.

\begin{figure}[tb]
    \centering
    \includegraphics[width=\textwidth]{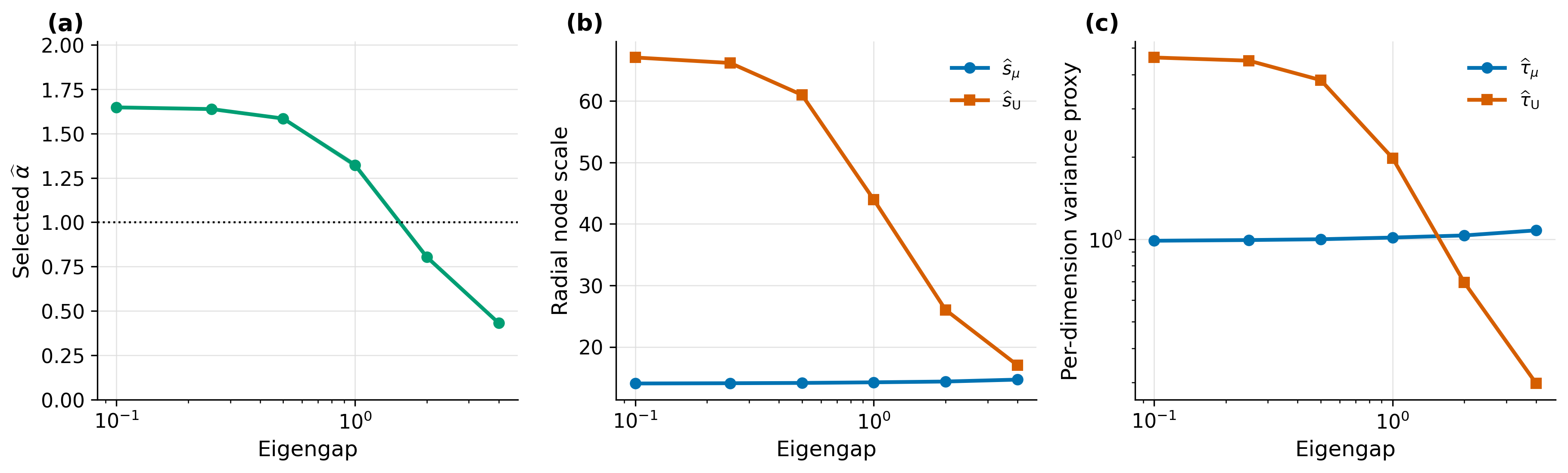}
    \caption{
    Scale diagnostics.  Panels show (a) the calibrated scale \(\widehat\alpha_{\rm rPCA}\), (b) the robust radial node scales for the mean and subspace components, and (c) per-dimension variance proxies.  
    }
    \label{fig:sim1-scale-diagnostics}
\end{figure}

Figure~\ref{fig:sim1-scale-diagnostics} shows the calibration mechanism.  The selected scale decreases from about \(1.65\) at \(\Delta=0.1\) to about \(0.43\) at \(\Delta=4\).  This agrees with the influence expansion: small eigengaps produce large Grassmann node errors, so the rule moves \(\alpha\) toward two and reduces the weight \(2-\alpha\) assigned to the subspace component.  As the eigengap grows, the Grassmann node scale decreases and the subspace receives more relative weight.  In this clean centered setting, fixed-scale and calibrated MoM have similar point-estimation errors; the main role of \(\widehat\alpha_{\rm rPCA}\) is to record the relative node-level uncertainty and determine the covariance geometry for inference.

\subsection{Factor-specific bad nodes and perturbation severity}
\label{subsec:sim-badnodes}

The second simulation studies robustness when a minority of nodes are corrupted.  We used the same spiked covariance model as in the previous experiment, with \(p=200\), \(r=5\), \(n=40000\), \(K=80\), and eigengap \(\Delta=1\).  Results were averaged over 50 Monte Carlo replications.  We considered two types of bad nodes.  In the mean-only contamination setting, the node mean was shifted while the node subspace was left unchanged.  In the subspace-only contamination setting, the node subspace was perturbed while the node mean was left unchanged.  We compared projector averaging, random subset PCA, fixed-scale MoM PCA with \(\alpha\in\{0.25,1,1.75\}\), and the robust scale-calibrated MoM estimator.

According to the factorwise robustness theory, the deterministic breakdown threshold of the median aggregation is governed by the fraction of bad nodes and is not changed by \(\alpha\).  However, the factorwise influence constants depend on scale:
\[
    \|\psi_{\alpha,\mu}\|\le \alpha^{-1/2},
    \qquad
    \|\psi_{\alpha,\mathcal U}\|\le (2-\alpha)^{-1/2}.
\]
Thus, large \(\alpha\) protects the mean coordinate by making mean deviations more visible in the product distance, whereas small \(\alpha\) protects the subspace coordinate by making subspace deviations more visible.  The simulation therefore asks not only whether MoM is robust, but also which factor is protected by a given scale.

\begin{figure}[tb]
    \centering
    \includegraphics[width=\textwidth]{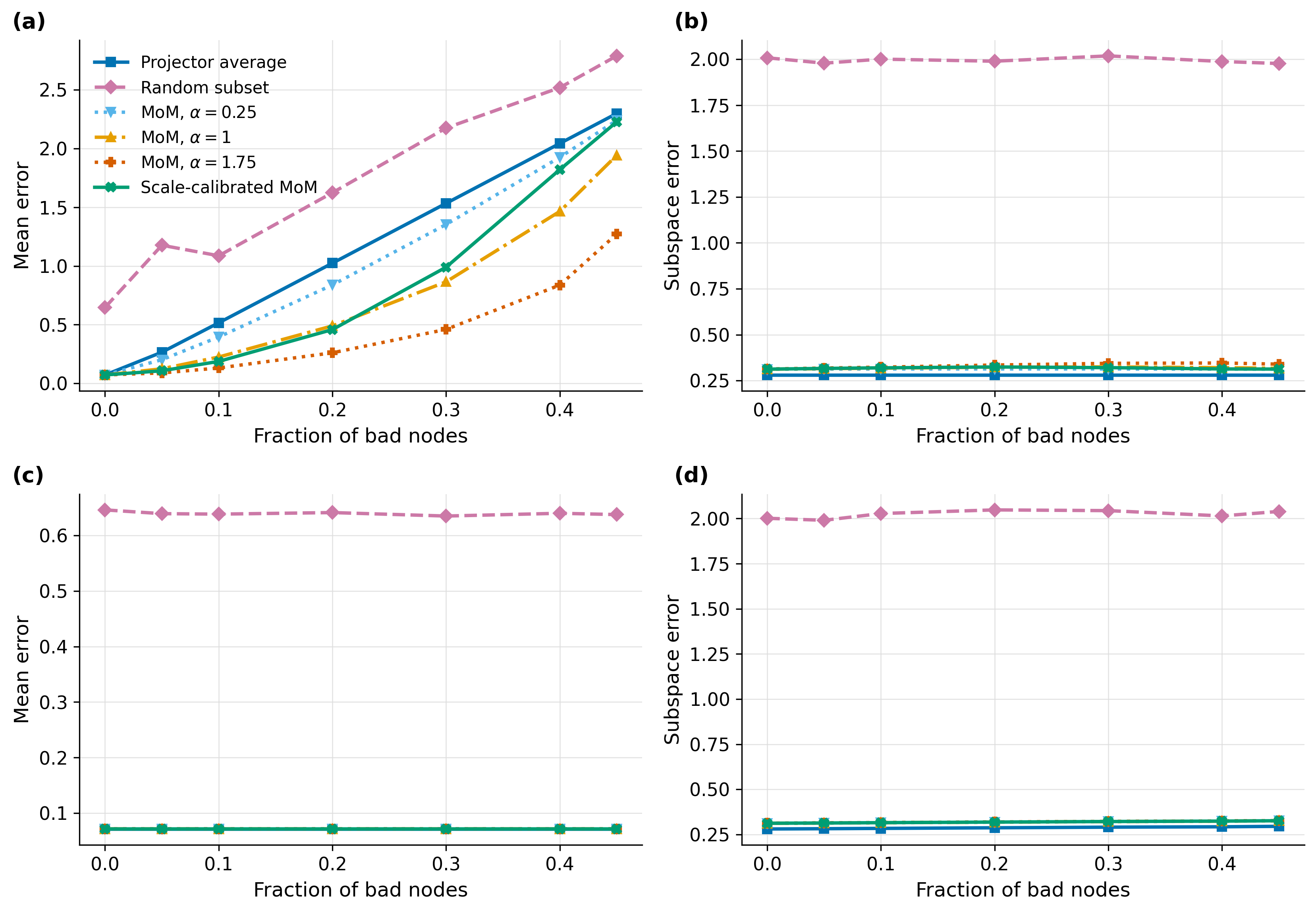}
    \caption{
Effect of increasing the fraction of corrupted nodes in the spiked PCA model with eigengap \(\Delta=1\).  Panels show (a) mean error under mean-only corruption, (b) subspace error under mean-only corruption, (c) mean error under subspace-only corruption, and (d) subspace error under subspace-only corruption.
}
    \label{fig:sim2-fraction}
\end{figure}

Figure~\ref{fig:sim2-fraction} reports the effect of increasing the fraction of bad nodes.  Under mean-only contamination, the mean error increases rapidly for projector averaging and random subset PCA.  The fixed-scale MoM estimators remain much more stable, and the ordering follows the influence calculation: \(\alpha=1.75\) gives the best protection of the mean, followed by \(\alpha=1\), while \(\alpha=0.25\) is least protective.  At \(40\%\) mean-only contamination, the average mean errors are approximately \(2.04\) for projector averaging, \(2.52\) for random subset PCA, \(1.46\) for fixed-scale MoM with \(\alpha=1\), \(0.84\) for fixed-scale MoM with \(\alpha=1.75\), and \(1.82\) for scale-calibrated MoM.  The calibrated estimator is therefore more stable than the nonrobust distributed baselines, but not the most protective for a purely mean-directed attack.

The calibrated rule estimates relative node dispersion, so mean-only bad nodes
can reduce the selected scale and thereby downweight the Euclidean component.
This protects the subspace but does not give the strongest possible mean
protection.  Conversely, fixed large \(\alpha\) is best for a known mean-only
attack, while smaller \(\alpha\) is preferable when subspace contamination is
the concern.  Thus the fixed-scale curves are not competitors only. They are
diagnostics that reveal the factorwise robustness tradeoff.

\begin{figure}[tb]
    \centering
    \includegraphics[width=\textwidth]{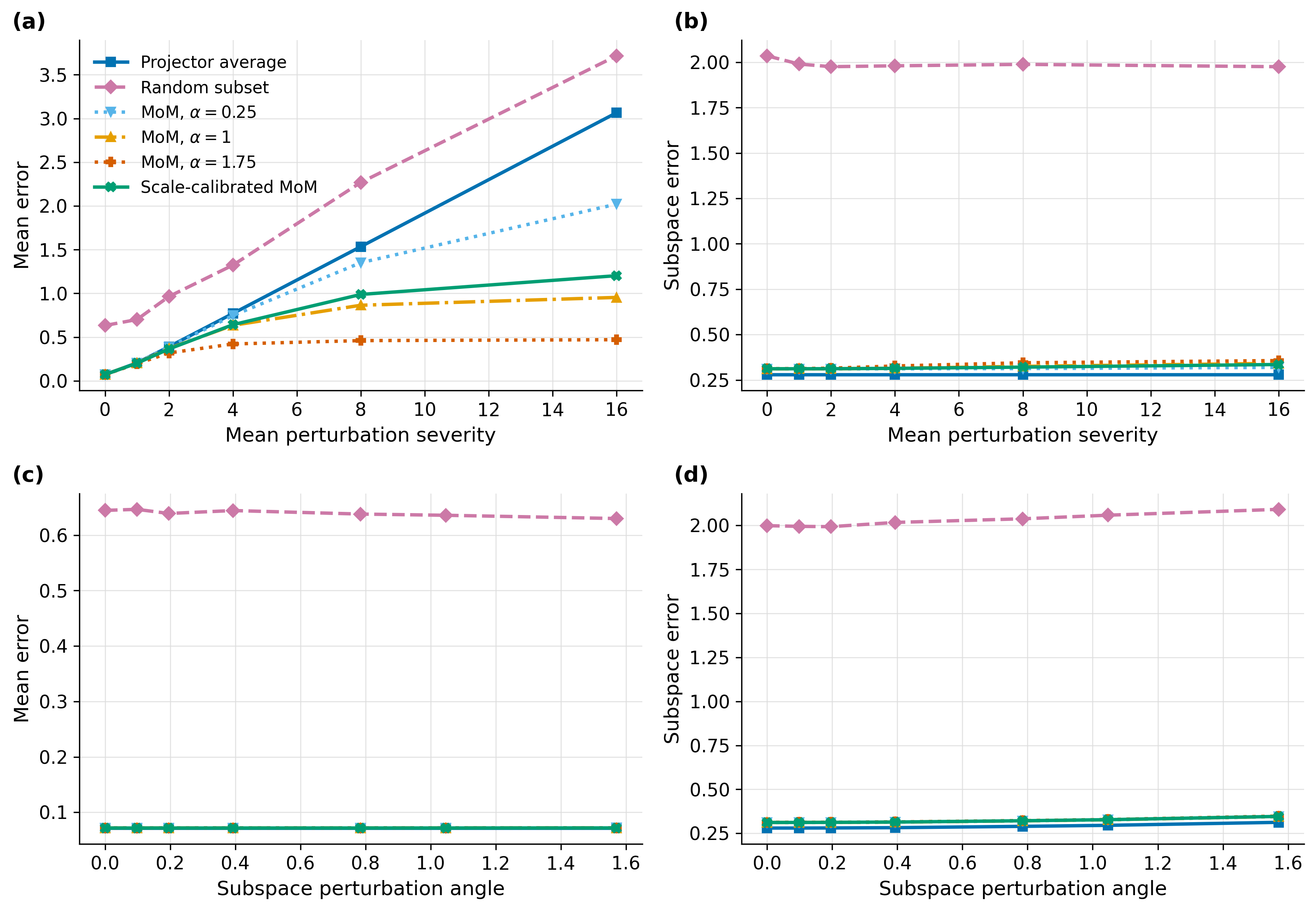}
    \caption{
Effect of increasing bad-node perturbation magnitude at a fixed contamination fraction. Panels show (a) mean error under increasing mean-shift severity, (b) subspace error under increasing mean-shift severity, (c) mean error under increasing subspace-tilt severity, and (d) subspace error under increasing subspace-tilt severity.
}
\label{fig:sim2-severity}
\end{figure}

Figure~\ref{fig:sim2-severity} studies perturbation magnitude at a fixed bad-node fraction.  For mean-only bad nodes, increasing the shift severity produces approximately monotone growth in mean error for all methods, but the rates differ substantially.  At the largest shift severity, projector averaging has mean error about \(3.07\), random subset PCA about \(3.72\), fixed-scale MoM with \(\alpha=1\) about \(0.95\), fixed-scale MoM with \(\alpha=1.75\) about \(0.47\), and scale-calibrated MoM about \(1.20\).  Thus, MoM aggregation substantially improves over the nonrobust baselines, and the fixed-\(\alpha\) curves confirm the predicted scale dependence of mean robustness.  For subspace-only perturbations, the subspace error increases with the perturbation angle, while the mean coordinate remains stable.  The calibrated and fixed-scale MoM estimates remain bounded throughout the severity range, whereas random subset PCA is again far less reliable.

Taken together, these simulations show that scale calibration should not be interpreted as a universally loss-minimizing choice under every adversarial pattern.  Rather, it is an adaptive inferential rule that balances mean and subspace uncertainty using node-level dispersion.  The fixed-scale sensitivity curves are therefore an important companion to the calibrated estimator: they reveal which factor is being protected by a given scale and explain why different contamination mechanisms favor different choices of \(\alpha\).

\subsection{Mouse brain single-cell data}
\label{subsec:real-mouse}

We next evaluate the proposed method on the 10x Genomics 1.3 million mouse brain single-cell RNA-seq dataset.  The data consist of expression profiles from embryonic mouse brain tissue, including cortex, hippocampus, and subventricular zone.  This example is well suited to our setting for two reasons.  First, PCA is a standard dimension-reduction step in single-cell analysis.  Second, the scale of the data makes distributed computation meaningful: after preprocessing, the analysis involves more than one million cells and thousands of genes.  We therefore use this example to study whether robust distributed PCA can recover stable mean--subspace summaries from node-level estimates.

We selected \(p=2000\) highly variable genes from the log-normalized expression matrix and used the full-data PCA solution as a reference.  The full-data estimator is not treated as an oracle truth; rather, it serves as a stable centralized benchmark against which distributed summaries can be compared.  We randomly split the cells into \(K\in\{50,100,200\}\) nodes and considered PCA ranks \(r\in\{10,20\}\).  For each node \(k\), we computed the node mean \(\widehat\mu_k\) and leading principal subspace \(\widehat{\mathcal U}_k\in\Gr(r,p)\).  We compared projector averaging, random subset PCA using one node-sized subsample, fixed-scale MoM PCA with \(\alpha=1\), and the proposed scale-calibrated MoM PCA.  The random splitting was repeated 10 times for each pair \((K,r)\).  Mean errors are measured by Euclidean distance to the full-data mean, and subspace errors are measured by Grassmann distance to the full-data principal subspace.

For a qualitative comparison, we also constructed two-dimensional PCA embeddings from the same node-level estimates.  The visualization uses a common subsample of cells.  Full-data PCA is used as the reference embedding.  For each competing distributed method, we centered its two-dimensional scores and aligned them to the full-data PCA scores by orthogonal Procrustes rotation.  This alignment removes arbitrary rotations and sign changes, so differences across panels reflect changes in the recovered embedding rather than coordinate conventions.  Because curated cell-type labels are not available for this dataset, colors are pseudo-labels obtained from a standard full-data PCA workflow followed by a \(k\)-nearest-neighbor graph with $k=15$ and Leiden clustering.  These pseudo-labels are used only to provide a common visual reference across methods.

\begin{figure}[tb]
    \centering
    \includegraphics[width=\textwidth]{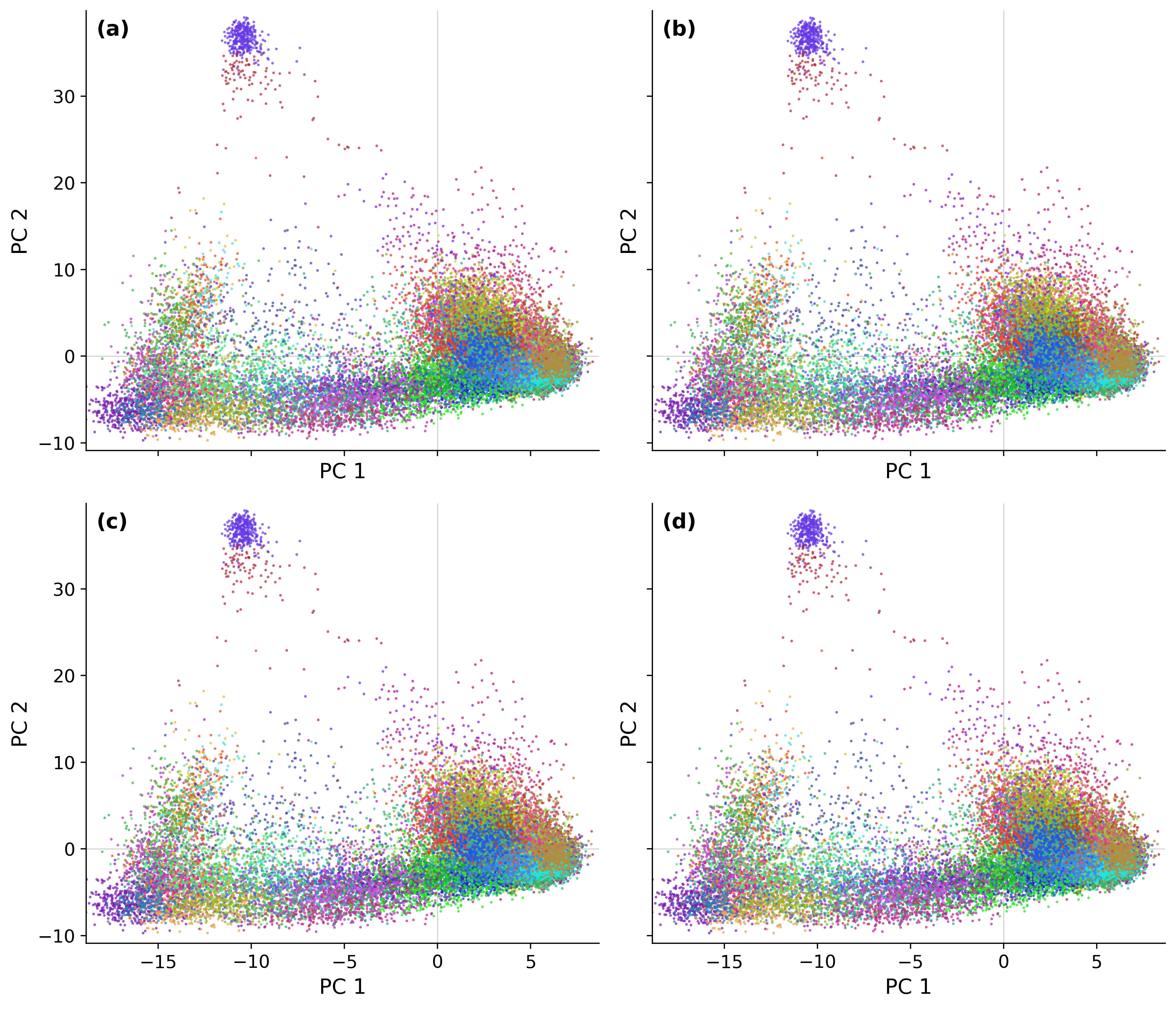}
    \caption{
    {10x mouse brain aligned PCA embeddings.}
    Two-dimensional embeddings of a common subsample of cells under four PCA summaries.  Full-data PCA is used as the reference, and all other embeddings are centered and aligned to it by orthogonal Procrustes matching.  Colors indicate pseudo-labels obtained from the full-data PCA \(k\)-nearest-neighbor and Leiden clustering pipeline; they are used only as a common visualization aid.  Panels show (a) full-data PCA, (b) projector-averaged distributed PCA, (c) fixed-scale MoM PCA with \(\alpha=1\), and (d) scale-calibrated MoM PCA.
    }
    \label{fig:real-mouse-embedding}
\end{figure}

\begin{figure}[tb]
    \centering
    \includegraphics[width=\textwidth]{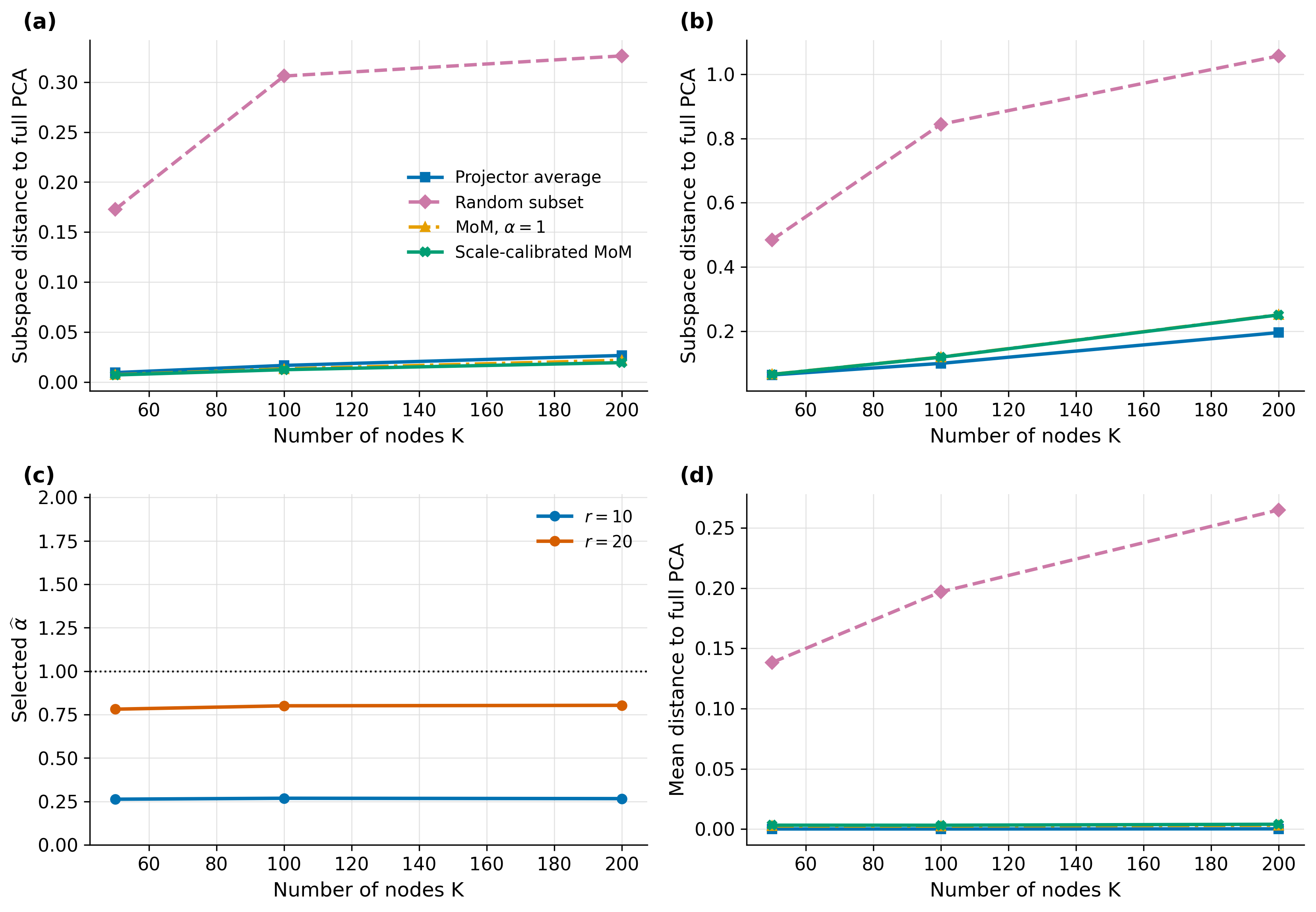}
    \caption{
    {10x mouse brain clean splitting.}
    Distributed PCA comparison using \(p=2000\) highly variable genes, \(K\in\{50,100,200\}\) nodes, and ranks \(r\in\{10,20\}\).  Full-data PCA is used as the centralized reference.  Random subset PCA uses one node-sized subsample, whereas projector averaging and the two MoM procedures aggregate all node estimates.  Panels show (a) subspace distance for \(r=10\), (b) subspace distance for \(r=20\), (c) selected calibrated scale, and (d) mean distance for \(r=10\).
    }
    \label{fig:real-mouse-clean}
\end{figure}

The embedding comparison in Figure~\ref{fig:real-mouse-embedding} shows that the distributed aggregation methods preserve the large-scale organization of the full-data PCA embedding.  The quantitative summaries in Figure~\ref{fig:real-mouse-clean} and Table~\ref{tab:real-mouse-selected} measure how closely the estimated mean--subspace pairs match the centralized reference.  Random subset PCA is unstable for both ranks and worsens as \(K\) increases.  Methods that aggregate all nodes are much closer to the full-data reference.  For \(r=10\), scale-calibrated MoM has the smallest average subspace distance among the distributed methods; for \(K=100\), the subspace errors are \(0.0167\) for projector averaging, \(0.0132\) for fixed-scale MoM, \(0.0124\) for scale-calibrated MoM, and \(0.3061\) for random subset PCA.

\begin{table}[tb]
\centering
\caption{
{Selected 10x mouse brain results.}
Mean and subspace errors are reported relative to the full-data PCA reference for \(K=100\).  Projector averaging is the standard distributed baseline, random subset PCA uses one node-sized subsample, and the MoM estimators aggregate node mean--subspace pairs on \(\R^p\times\Gr(r,p)\).  
}
\label{tab:real-mouse-selected}
\begin{tabular}{lcccc}
\toprule\noalign{}
Rank & Method & Mean error & Subspace error & \(\widehat\alpha\) \\
\midrule
\multirow{4}{*}{\(r=10\)}
& Projector average & \(0.00009\) & \(0.01667\) & -- \\
& Random subset & \(0.19687\) & \(0.30608\) & -- \\
& MoM, \(\alpha=1\) & \(0.00253\) & \(0.01319\) & \(1.000\) \\
& Scale-calibrated MoM & \(0.00320\) & \(0.01239\) & \(0.269\) \\
\midrule
\multirow{4}{*}{\(r=20\)}
& Projector average & \(0.00009\) & \(0.09960\) & -- \\
& Random subset & \(0.20008\) & \(0.84422\) & -- \\
& MoM, \(\alpha=1\) & \(0.00583\) & \(0.11932\) & \(1.000\) \\
& Scale-calibrated MoM & \(0.00593\) & \(0.11884\) & \(0.801\) \\
\bottomrule
\end{tabular}
\end{table}

Table~\ref{tab:real-mouse-selected} reports representative numerical values for \(K=100\).  The improvement of scale-calibrated MoM over fixed-scale MoM is modest but systematic for \(r=10\) that it reduces subspace discrepancy while incurring only a small increase in mean error. The behavior at \(r=20\) is different.  Projector averaging has the smallest subspace distance to the full-data reference, while the two MoM procedures remain far more stable than random subset PCA.  The selected scale also changes with rank.  For \(r=10\), \(\widehat\alpha_{\rm rPCA}\) is close to \(0.27\), whereas for \(r=20\), it is close to \(0.8\).  This indicates that the relative node-level mean and subspace uncertainties depend on the PCA resolution.  Thus, the calibrated scale is not merely a numerical tuning parameter but reflects the empirical balance between Euclidean and Grassmann uncertainty at the node level.

The mean-error panel in Figure~\ref{fig:real-mouse-clean} shows the complementary side of this tradeoff.  Projector averaging has the smallest mean discrepancy in the uncorrupted setting, as expected from an averaging procedure.  The MoM estimators have slightly larger, but still small, mean errors.  This is consistent with the role of robust aggregation: the median is not designed to improve on averaging under clean homogeneous data.  Instead, it provides a stable distributed summary whose geometry is calibrated to observed node-level uncertainty.  The clean-data results therefore show that scale-calibrated MoM remains competitive with standard distributed PCA while avoiding the instability of relying on a single random subset.

We also performed a node-level stress test at \(K=50\) and \(r=10\).  The goal is not to model a specific biological artifact, but to probe the factorwise robustness mechanism predicted by the theory.  We considered two artificial corruption patterns.  In the mean-only setting, a fraction of node means was shifted while the corresponding node subspaces were left unchanged.  In the subspace-only setting, a fraction of node subspaces was perturbed while the corresponding node means were left unchanged.

\begin{figure}[tb]
    \centering
    \includegraphics[width=\textwidth]{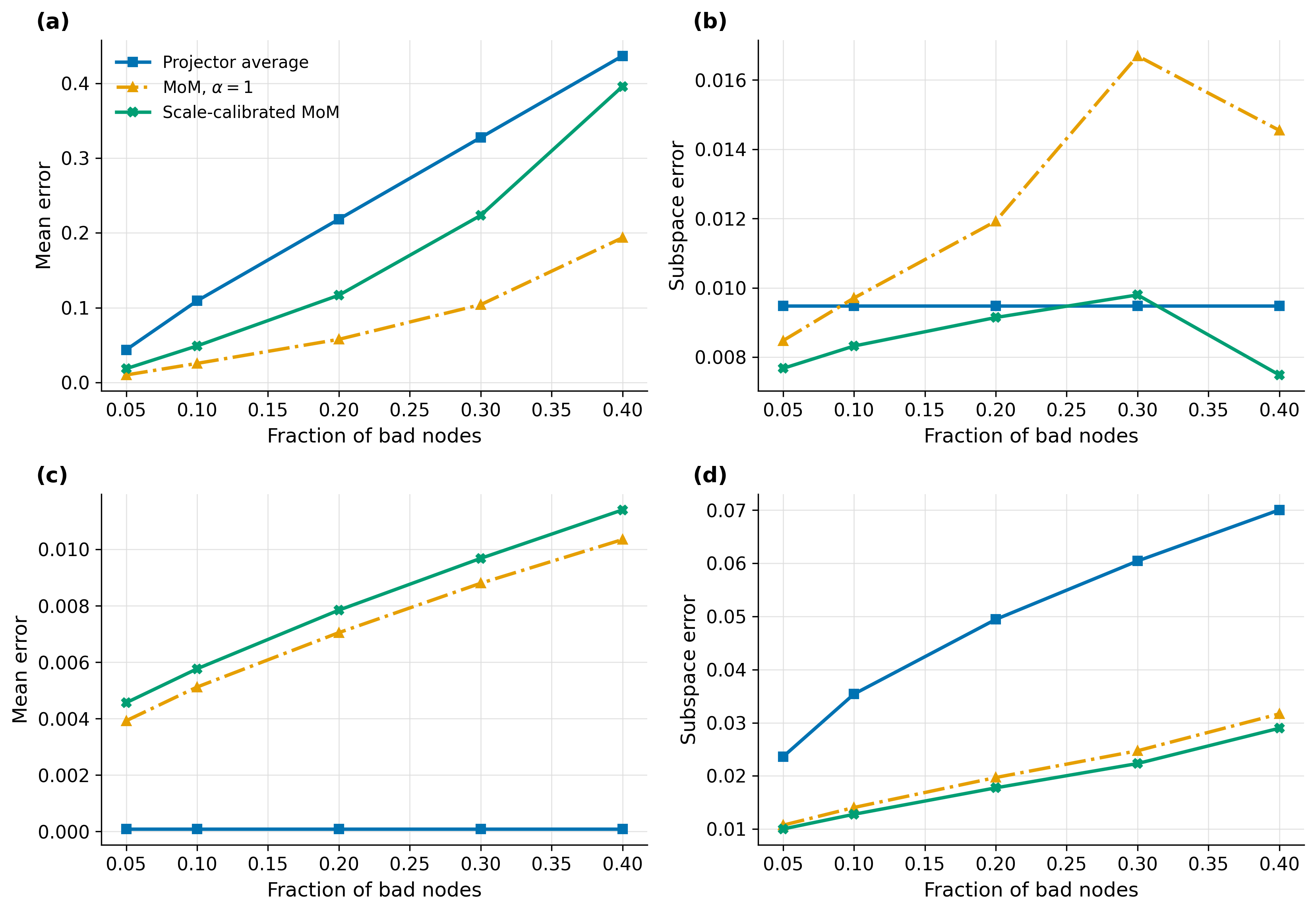}
    \caption{
    {10x mouse brain bad-node stress test.}
    Node-level corruption experiment with \(K=50\) and \(r=10\).  Panels show (a) mean error under mean-only corruption, (b) subspace error under mean-only corruption, (c) mean error under subspace-only corruption, and (d) subspace error under subspace-only corruption.
    }
    \label{fig:real-mouse-stress}
\end{figure}

Figure~\ref{fig:real-mouse-stress} shows the same mean--subspace tradeoff
observed in simulation.  Under mean-only corruption, fixed-scale MoM better
protects the mean, whereas the calibrated estimator better preserves the
subspace.  At \(40\%\) corrupted nodes, the calibrated estimator has larger mean
error than fixed-scale MoM but smaller subspace error.  Under subspace-only
corruption, scale-calibrated MoM gives the smallest subspace error among the
three methods.  These results reinforce that scale calibration learns from
node-level dispersion rather than uniformly minimizing every factorwise loss.

\section{Conclusion}
\label{sec:conclusion}

This paper studied robust distributed PCA through the geometry of
\(\R^p\times\Gr(r,p)\).  The main message is that median-of-means aggregation of
node PCA estimates is not only a robustness device, but also an inferential
design problem.  A node estimate contains a Euclidean mean and a Grassmann
principal subspace, and the relative scale assigned to these components
determines how node-level disagreement is measured.  Consequently, scale affects
covariance, finite-block bias, concentration, and factorwise robustness.

The theory explains why scale calibration is different in the MoM setting than
in ordinary product medians.  For ordinary product medians, changing factor
weights can change the population center.  In the present problem, the
aggregated objects are node estimators, so scale acts on the node-error
distribution. Since the mean influence is \(X-\mu_0\) and the subspace influence is an eigengap-weighted covariance perturbation, the relevant scale is a property of
node-level uncertainty rather than raw measurement units.

The empirical results support this interpretation.  The calibrated scale tracks
eigengap-driven changes in subspace variability, and the bad-node experiments
confirm the predicted factorwise tradeoff: large \(\alpha\) protects the mean,
whereas small \(\alpha\) protects the subspace.  The 10x mouse brain analysis
shows that the method is feasible at large scale and that the selected scale
changes with PCA rank.  The method is therefore best viewed as a robust
distributed alternative for heterogeneous, heavy-tailed, or partially corrupted
settings, not as a replacement for standard PCA in clean centralized data.

Several directions remain open.  The present theory is fixed-dimensional, while
many PCA applications involve growing \(p\), growing \(r\), or structural
constraints such as sparsity.  Robust node-level covariance estimators could be
incorporated when individual nodes are themselves contaminated.  The joint design
of the number of nodes \(K\), block size \(b\), and scale \(\alpha\) also merits
further study.  More broadly, the same principle applies to any distributed
procedure that produces heterogeneous product-valued summaries: the final robust
aggregation should calibrate scale from node-level uncertainty rather than
impose it externally.

\bibliographystyle{dcu}
\bibliography{references}

\newpage  
\appendix

\section*{Appendix}\label{app:main}

\addcontentsline{toc}{section}{Appendix}

\renewcommand{\thesection}{\Alph{section}}

\setcounter{section}{0}

\section{Proofs}

In this section, all references to theorem-like statements and equations correspond to the numbering used in the main manuscript.

\subsection{Proof of Theorem~1}
The expansion of the mean is immediate:
\[
    \sqrt b(\widehat\mu_k-\mu_0)=b^{-1/2}\sum_{i\in B_k}(X_i-\mu_0).
\]
For the covariance, replacing \(\mu_0\) by \(\widehat\mu_k\) changes the covariance estimator by a term
\[
    (\widehat\mu_k-\mu_0)(\widehat\mu_k-\mu_0)^T,
\]
which is \(O_p(b^{-1})\) under the moment assumption and is therefore negligible after multiplication by \(\sqrt b\).  Thus
\[
    \sqrt b(\widehat\Sigma_k-\Sigma)
    =b^{-1/2}\sum_{i\in B_k}
    \{(X_i-\mu_0)(X_i-\mu_0)^T-\Sigma\}+o_p(1).
\]
Let \(P_0=U_rU_r^T\) be the spectral projector onto \(\U_0\), and let \(\widehat P_k\) be the projector onto \(\widehat\U_k\).  By the differentiability of spectral projectors under the eigengap condition,
\[
    \widehat P_k-P_0
    =\calL_\Sigma(\widehat\Sigma_k-\Sigma)+O_p(\|\widehat\Sigma_k-\Sigma\|^2),
\]
where
\[
    \calL_\Sigma(E)=
    \sum_{j=1}^r\sum_{a=1}^{p-r}
    {u_{r+a}^TEu_j\over \lambda_j-\lambda_{r+a}}
    (u_{r+a}u_j^T+u_ju_{r+a}^T).
\]
The moment assumption implies \(\|\widehat\Sigma_k-\Sigma\|=O_p(b^{-1/2})\), so the quadratic remainder is \(o_p(b^{-1/2})\).  In the Grassmann normal chart at \(\U_0\), the tangent coordinate matrix for \(\widehat\U_k\) is, to first order, the off-diagonal block
\[
    U_\perp^T\calL_\Sigma(\widehat\Sigma_k-\Sigma)U_r.
\]
Substituting the covariance expansion yields
\[
    \sqrt b\log_{\U_0}(\widehat\U_k)
    =b^{-1/2}\sum_{i\in B_k}B(X_i)+o_p(1),
\]
with entries defined in (15).  Combining the mean and subspace components gives (16).  The multivariate central limit theorem applies because \(\E\|X-\mu_0\|^{4+\delta}<\infty\) implies \(\zeta(X)\) has finite second moment.  The uniform remainder condition (19) is stated explicitly for growing \(K\).

\subsection{Proof of Theorem~2}
Work on the high-probability event that all node subspaces lie in a common Grassmann normal neighborhood of \(\U_0\).  For bounded tangent vectors \(v,w\), the squared product distance between \(\exp_{\theta_0}(v/\sqrt b)\) and \(\exp_{\theta_0}(w/\sqrt b)\) has the local expansion
\[
    d_\alpha\{\exp_{\theta_0}(v/\sqrt b),\exp_{\theta_0}(w/\sqrt b)\}^2
    =b^{-1}\|v-w\|_{H_\alpha}^2+O(b^{-2}(1+\|v\|+\|w\|)^4),
\]
where the remainder is uniform over \(\alpha\in I_\eps\).  Taking square roots gives
\[
    d_\alpha\{\exp_{\theta_0}(v/\sqrt b),\exp_{\theta_0}(w/\sqrt b)\}
    =b^{-1/2}\|v-w\|_{H_\alpha}+O(b^{-3/2}(1+\|v\|+\|w\|)^3).
\]
Set \(w=W_{k,b}\).  Multiplying the median objective by \(\sqrt b\) yields
\[
    \sqrt b\,Q_{K,\alpha}(\exp_{\theta_0}(v/\sqrt b))
    ={1\over K}\sum_{k=1}^K\|v-W_{k,b}\|_{H_\alpha}+\Delta_{K,b,\alpha}(v),
\]
where \(\sup_{v\in C,\alpha\in I_\eps}|\Delta_{K,b,\alpha}(v)|=o_p(1)\) for fixed \(K\) on every compact set \(C\).  The objectives are convex in the tangent approximation and locally coercive under Assumption (A2).  Standard argmin transfer therefore gives (21).  For growing \(K\), the same argument applies at the \(K^{-1/2}\) scale under the stated stronger approximation condition.  The sufficient condition \(\sqrt K/b\to0\) follows from the displayed remainder bound and a bounded third moment for \(W_{k,b}\).

\subsection{Proof of Theorem~3}
By Theorem ~1, \((W_{1,b},\ldots,W_{K,b})\dto(W_1,
\ldots,W_K)\).  The map sending \((w_1,
\ldots,w_K)\) to the unique minimizer of \(K^{-1}\sum_k\|s-w_k\|_{H_\alpha}\) is continuous at configurations with a unique median.  Hence the continuous mapping theorem gives \(\widehat s_{K,b,\alpha}\dto T_{\alpha,K}\).  The local reduction theorem transfers this convergence to \(\sqrt b\log_{\theta_0}\widetilde\theta_{n,\alpha}\).  Multiplication by \(\sqrt K\) gives the root-\(n\) form.

\subsection{Proof of Theorem~4}
Transform the spatial median problem by writing \(y=H_\alpha^{1/2}s\) and \(Y_{k,b}=H_\alpha^{1/2}W_{k,b}\).  The median of \(W_{k,b}\) under \(\|\cdot\|_{H_\alpha}\) is \(s_{\alpha,b}=H_\alpha^{-1/2}y_{\alpha,b}\), where \(y_{\alpha,b}\) is the ordinary Euclidean spatial median of \(Y_b\).  The estimating equation in \(y\)-coordinates is
\[
    {1\over K}\sum_{k=1}^K {Y_{k,b}-y\over\|Y_{k,b}-y\|}=0.
\]
At \(y=y_{\alpha,b}\), the summand has mean zero.  The derivative of the population score is \(-A_{\alpha,b}\), where \(A_{\alpha,b}\) is defined in (13).  A Taylor expansion of the estimating equation gives
\[
    \sqrt K(\widehat y_{K,b,\alpha}-y_{\alpha,b})
    =A_{\alpha,b}^{-1}{1\over\sqrt K}\sum_{k=1}^K U_{k,\alpha,b}+o_p(1),
\]
where \(U_{k,\alpha,b}=Y_{k,b}-y_{\alpha,b}\over \|Y_{k,b}-y_{\alpha,b}\|\).  The multivariate triangular-array CLT gives a normal limit with covariance \(A_{\alpha,b}^{-1}S_{\alpha,b}A_{\alpha,b}^{-T}\).  Transforming back by \(H_\alpha^{-1/2}\) gives (28).  Finally, by Theorem~2,
\[
    \sqrt n\log_{\theta_0}\widetilde\theta_{n,\alpha}
    =\sqrt K\widehat s_{K,b,\alpha}+o_p(1),
\]
which yields (29) and the stated limits.

\subsection{Proof of Corollary~5}
If \(W_b\stackrel d=-W_b\), then \(s=0\) minimizes \(\E\|s-W_b\|_{H_\alpha}\) by symmetry and uniqueness, so \(s_{\alpha,b}=0\).  The second statement follows by substituting (14) into \(\sqrt K s_{\alpha,b}\).

\subsection{Proof of Theorem~6}
The proof is the limiting version of the argument in Theorem~4.  With \(Y_\alpha=H_\alpha^{1/2}W\), the Euclidean spatial median covariance in transformed coordinates is \(A_\alpha^{-1}S_\alpha A_\alpha^{-T}\).  Since \(s=H_\alpha^{-1/2}y\), the covariance in the original product tangent coordinates is obtained by pre- and post-multiplying by \(H_\alpha^{-1/2}\), giving (27).

\subsection{Proof of Proposition~7}
If \(Y_\alpha=RU\), where \(U\) is uniform on the unit sphere in \(\R^d\) and independent of \(R/\sigma\sim\chi_d\), then \(S_\alpha=\Var(U)=I_d/d\).  Also
\[
    A_\alpha=\E\left[{I-UU^T\over R}\right]
    =\E(R^{-1})\left(I_d-{I_d\over d}\right)
    ={d-1\over d}\E(R^{-1})I_d.
\]
Thus
\[
    A_\alpha^{-1}S_\alpha A_\alpha^{-T}
    ={d\over(d-1)^2}\{\E(R^{-1})\}^{-2}I_d.
\]
Because \(R=\sigma\chi_d\) and \(\E(\chi_d^{-1})=2^{-1/2}\Gamma((d-1)/2)/\Gamma(d/2)\), the constant is (38).  Stirling's formula gives \(c_d\to1\).

\subsection{Proof of Theorem~8}
The map \((x,y)\mapsto 2\{y^2/[r(p-r)]\}/\{x^2/p+y^2/[r(p-r)]\}\) is continuous on \((0,\infty)^2\).  The first statement follows from the continuous mapping theorem.  Under block-isotropy, radial median scales are proportional to the corresponding standard deviations, with constants depending only on the block dimensions.  The dimension normalization removes those constants when the blocks are calibrated per tangent dimension, yielding the whitening scale up to the stated known constants.

\subsection{Proof of Theorem~9}
Uniform convergence of \(\widehat V_\alpha\) implies uniform convergence of \(\widehat R(\alpha)\) to \(R(\alpha)\) on the compact set \(I_\eps\).  Consistency follows from the standard argmin theorem.  For the second statement, the first-order condition \(\widehat R'(\widehat\alpha^*)=0\) and a Taylor expansion around \(\alpha^*\) give
\[
    \sqrt K(\widehat\alpha^*-\alpha^*)
    =-{\sqrt K\{\widehat R'(\alpha^*)-R'(\alpha^*)\}\over R''(\alpha^*)}+o_p(1),
\]
which completes the proof.

\subsection{Proof of Theorem~10}
The stochastic equicontinuity condition gives
\[
    \sqrt K\{\widehat s_{K,b,\widehat\alpha}-s_{\widehat\alpha,b}
    -\widehat s_{K,b,\alpha_0}+s_{\alpha_0,b}\}=o_p(1).
\]
If \(s_{\alpha,b}=0\) for all \(\alpha\), this implies that \(\sqrt K\widehat s_{K,b,\widehat\alpha}\) has the same limit as \(\sqrt K\widehat s_{K,b,\alpha_0}\), and the local reduction transfers the result to \(\sqrt n\log_{\theta_0}\widetilde\theta_{n,\widehat\alpha}\).  If the path is nonzero, the difference is exactly the additional term \(\sqrt K\{s_{\widehat\alpha,b}-s_{\alpha_0,b}\}\) plus the negligible stochastic equicontinuity remainder.  Under the stated bias expansion and \(\widehat\alpha-\alpha_0=O_p(K^{-1/2})\), continuity of \(a_\alpha\) gives a term of order \(O_p(b^{-1/2})\).

\subsection{Proof of Theorem~11}
Let \(G\) be the set of good nodes and \(B\) its complement.  Let \(R=d_\alpha(\widetilde z_\alpha,\theta_0)\).  Since \(\widetilde z_\alpha\) minimizes the sum of distances,
\[
    \sum_{k=1}^K d_\alpha(\widetilde z_\alpha,z_k)
    \le \sum_{k=1}^K d_\alpha(\theta_0,z_k).
\]
For \(k\in G\), the triangle inequality gives
\[
    d_\alpha(\widetilde z_\alpha,z_k)-d_\alpha(\theta_0,z_k)
    \ge R-2r.
\]
For \(k\in B\), it gives
\[
    d_\alpha(\widetilde z_\alpha,z_k)-d_\alpha(\theta_0,z_k)
    \ge -R.
\]
Therefore
\[
    0\ge |G|(R-2r)-|B|R=(|G|-|B|)R-2|G|r.
\]
Because \(|G|/K\ge1/2+\gamma\), \(|G|-|B|\ge2\gamma K\) and \(|G|\le K\).  Hence \(R\le (1+1/(2\gamma))r\).

\subsection{Proof of Theorem~12}
Let \(C_\alpha=\alpha v_\mu+(2-\alpha)v_\U\).  By Markov's inequality,
\[
    \Pbb\{d_\alpha(\widehat\theta_k,\theta_0)>2\sqrt{C_\alpha/b}\}\le {1\over4}.
\]
Thus a node is good with probability at least \(3/4\).  Hoeffding's inequality implies that, with probability at least \(1-\exp(-c_2K)\), at least \((1/2+\gamma)K\) nodes are good for a universal \(\gamma>0\), for example \(\gamma=1/8\).  Applying Theorem 11 with \(r=2\sqrt{C_\alpha/b}\) gives the result with a universal constant \(c_1\).

\subsection{Proof of Corollary~13}
The scaled distance satisfies
\[
    d_\alpha(\theta,\theta_0)^2
    =\alpha\|\mu-\mu_0\|^2+(2-\alpha)d_{\Gr}(\U,\U_0)^2.
\]
Therefore \(\|\mu-\mu_0\|\le d_\alpha(\theta,\theta_0)/\sqrt\alpha\) and \(d_{\Gr}(\U,\U_0)\le d_\alpha(\theta,\theta_0)/\sqrt{2-\alpha}\).  Combine these inequalities with Theorem~12.

\subsection{Proof of Theorem~14}
The boundedness statement follows from the same deterministic good-node argument as Theorem 11.  For the influence constants, write
\[
    \|\psi_{\alpha,\mu}(w)\|
    ={\|w_\mu\|\over\{\alpha\|w_\mu\|^2+(2-\alpha)\|w_\U\|^2\}^{1/2}}
    \le \alpha^{-1/2}.
\]
Equality is approached when \(w_\U=0\) and \(w_\mu\ne0\).  The subspace component is identical, giving \((2-\alpha)^{-1/2}\).

\subsection{Proof of Theorem~15}
By the local reduction theorem, the bootstrap statistic is asymptotically equivalent to the bootstrap spatial median of the empirical node errors.  Conditional on the observed nodes, the resampled node errors are i.i.d. from the empirical distribution.  The spatial median is a smooth \(Z\)-estimator under Assumption (A2), and the empirical derivative and score covariance converge in probability to their population limits.  The standard bootstrap theorem for smooth finite-dimensional \(Z\)-estimators therefore yields
\[
    \sqrt K(\widehat s_{K,b,\alpha}^*-\widehat s_{K,b,\alpha})\dto \Normal(0,V_\alpha)
\]
conditionally in probability.  Multiplication by the local factor relating \(\sqrt K\) in tangent error space to \(\sqrt n\) on the original manifold gives (58).  If scale is estimated and its first-order effect is not negligible, the bootstrap must reproduce the same scale-selection step to mimic the full estimator.

\section{Additional simulation experiments}
\label{sec:supp-simulations}

This section reports two additional simulation experiments that support the main theoretical claims.  The simulations in the main paper focus on eigengap-driven scale calibration and factor-specific bad nodes.  Here we isolate two more technical aspects of the theory.  The first supplementary experiment validates the explicit covariance formula \(V_\alpha\) in an oracle Gaussian node-error setting, where the node errors are generated directly and no PCA estimation error is present.  The second supplementary experiment separates the fixed-\(K\), growing-\(K\), and aggressive-\(K\) regimes, illustrating the finite-block bias phenomenon described in Theorem~4 and Corollary~5.

\subsection{Oracle Gaussian node-error experiment}
\label{subsec:supp-oracle-covariance}

The covariance formula in Theorem~6 is derived for the spatial median of scaled node influence errors.  To check this formula without confounding from finite-sample PCA estimation, we first simulate node errors directly from a Gaussian tangent-space model.  Specifically, we generate
\[
    W_k\sim N(0,\Gamma),
\]
where the covariance \(\Gamma\) has separate mean and subspace blocks.  We vary the mean-to-subspace variance ratio over
\[
    0.25,\ 0.5,\ 1,\ 2,\ 4,
\]
and use two cross-block correlation settings,
\[
    \rho\in\{0,0.3\}.
\]
For each configuration and each value of \(\alpha\), we compute the theoretical covariance trace from the formula
\[
    V_\alpha
    =
    H_\alpha^{-1/2}
    A_\alpha^{-1}
    S_\alpha
    A_\alpha^{-T}
    H_\alpha^{-1/2},
\]
and compare it with the empirical covariance of replicated scaled spatial medians.

\begin{figure}[ht]
    \centering
    \includegraphics[width=\textwidth]{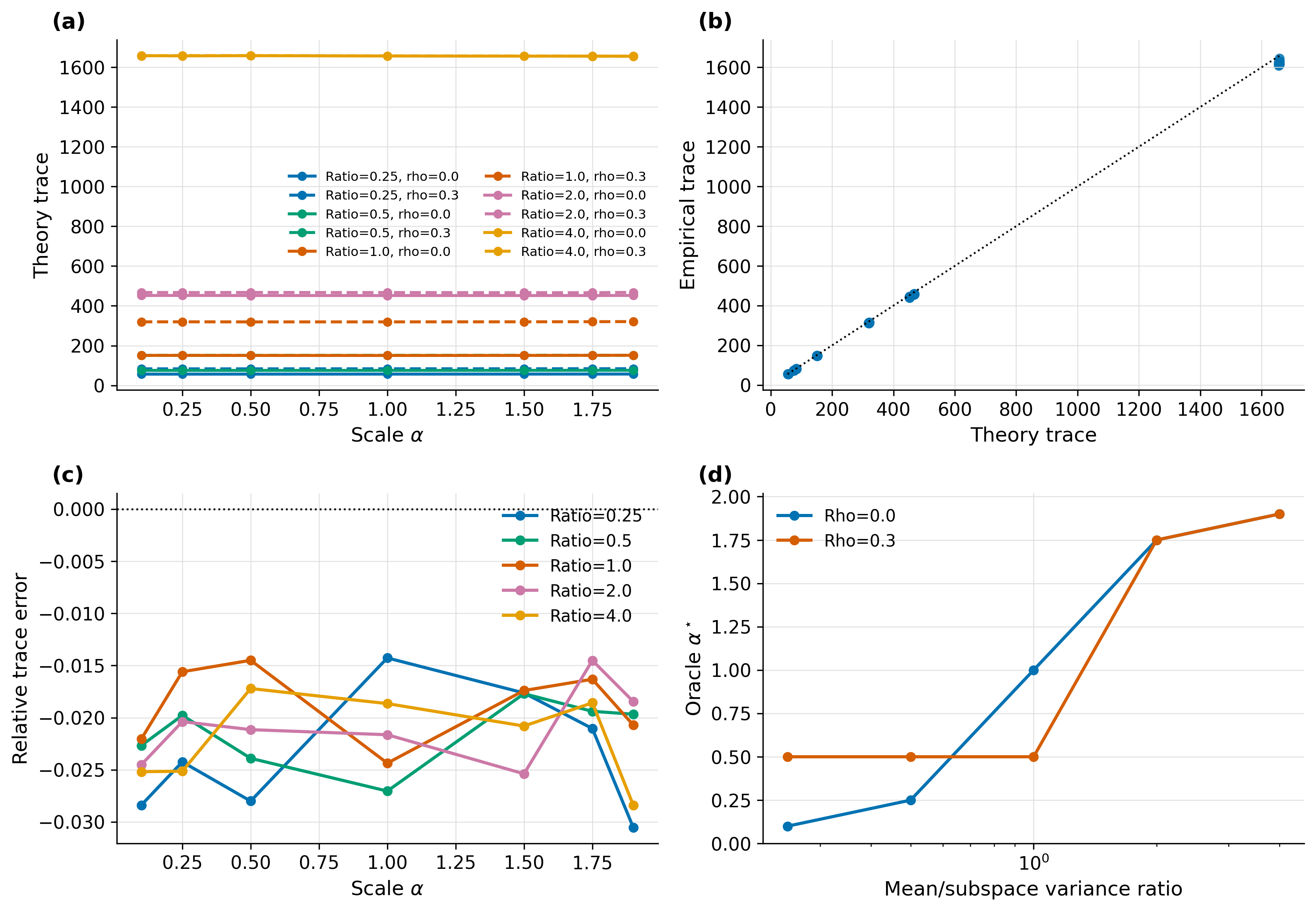}
    \caption{
    Oracle covariance check.
    Gaussian node errors are generated directly in tangent coordinates, so the comparison isolates the spatial-median covariance formula without additional PCA estimation error.  Panels show (a) theoretical covariance trace as a function of scale, (b) empirical trace versus theoretical trace, (c) relative trace error, and (d) oracle scale \(\alpha^*\) as a function of the mean-to-subspace variance ratio.
    }
    \label{fig:supp-s1-oracle}
\end{figure}

Figure~\ref{fig:supp-s1-oracle} summarizes the comparison.  The theoretical trace changes with the variance ratio and cross-block correlation, but is nearly flat in \(\alpha\) for the tested Gaussian configurations.  The empirical trace closely follows the theoretical trace: the scatter plot in panel (b) lies close to the diagonal, and the relative trace errors in panel (c) are small.  The oracle scale \(\alpha^*\), computed by minimizing the theoretical trace over the grid, changes with the mean--subspace variance ratio, as expected.  When mean variability is smaller relative to subspace variability, the optimal scale tends to put less weight on the subspace component; when mean variability is larger, the optimal scale moves in the opposite direction.

\begin{table}[ht]
\centering
\caption{
Oracle covariance validation.
Mean and maximum absolute relative errors between the theoretical covariance trace and the empirical covariance trace of replicated scaled spatial medians.  The ratio column gives the mean-to-subspace variance ratio, and \(\rho\) controls cross-block correlation.  The errors are small across all tested configurations.
}
\label{tab:supp-s1-error-summary}
\begin{tabular}{rrrrr}
\toprule
Ratio & \(\rho\) & Mean abs. rel. error & Max abs. rel. error & Mean norm \\
\midrule
0.25 & 0.0 & 0.0234 & 0.0305 & 0.1076 \\
0.25 & 0.3 & 0.0216 & 0.0309 & 0.1289 \\
0.50 & 0.0 & 0.0215 & 0.0270 & 0.1231 \\
0.50 & 0.3 & 0.0203 & 0.0253 & 0.1739 \\
1.00 & 0.0 & 0.0187 & 0.0244 & 0.1729 \\
1.00 & 0.3 & 0.0205 & 0.0264 & 0.2521 \\
2.00 & 0.0 & 0.0209 & 0.0254 & 0.2973 \\
2.00 & 0.3 & 0.0208 & 0.0244 & 0.3056 \\
4.00 & 0.0 & 0.0220 & 0.0284 & 0.5766 \\
4.00 & 0.3 & 0.0192 & 0.0264 & 0.5756 \\
\bottomrule
\end{tabular}
\end{table}

\begin{figure}[ht!]
    \centering
    \includegraphics[width=\textwidth]{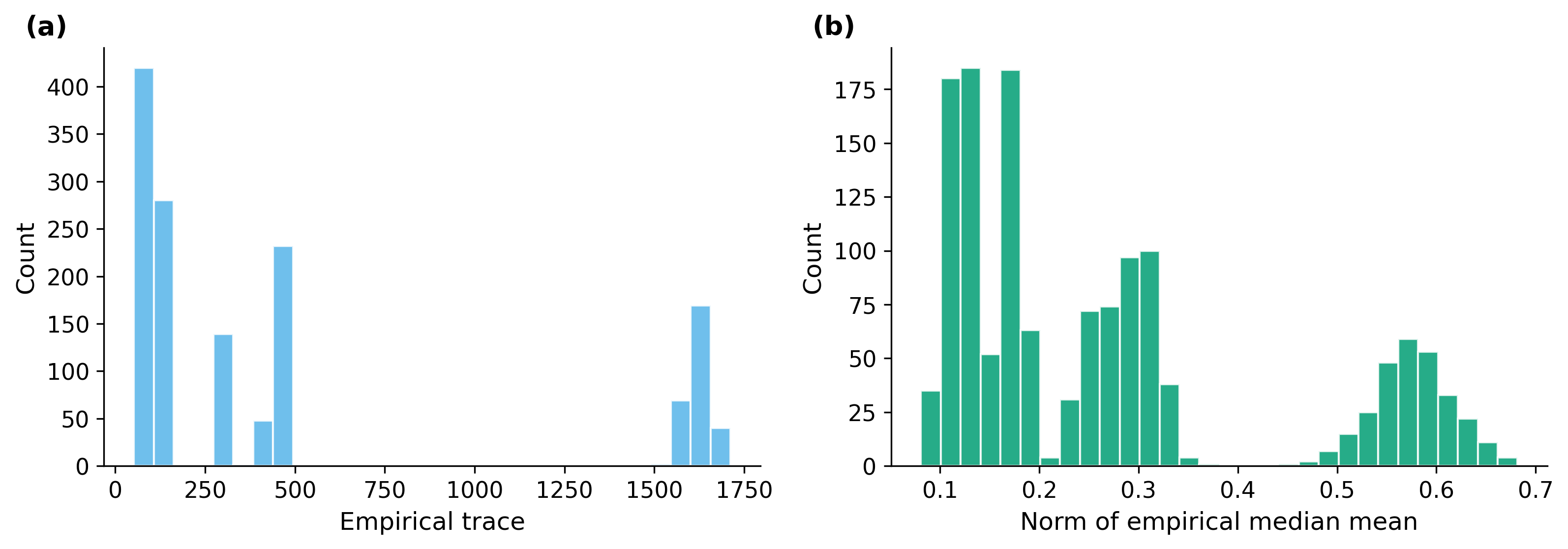}
    \caption{
    Oracle replication distributions.
    Distributional summaries from the Gaussian node-error covariance experiment.  Panel (a) shows the empirical covariance trace across replicated spatial medians, and panel (b) shows the norm of the empirical spatial-median mean.  The separated trace clusters correspond to different variance-ratio regimes, while the empirical median means remain small under centered Gaussian node errors.
    }
    \label{fig:supp-s1-replications}
\end{figure}

Table~\ref{tab:supp-s1-error-summary} gives a compact numerical summary.  Across all configurations, the mean absolute relative trace error is about \(2\%\), and the maximum absolute relative trace error is about \(3.1\%\).  Thus the empirical spatial-median covariance agrees well with the theoretical covariance formula.  Figure~\ref{fig:supp-s1-replications} further shows the distribution of empirical traces and the norm of the empirical median mean across replications.  The trace distribution reflects the different variance-ratio regimes, while the empirical median means remain small relative to the scale of the covariance trace, consistent with centered Gaussian node errors.

\subsection{Fixed-\(K\), growing-\(K\), and aggressive-\(K\) regimes}
\label{subsec:supp-asymptotic-regimes}

The second simulation examines the two asymptotic regimes.  The local reduction theorem shows that the product-manifold MoM estimator behaves like a spatial median of node errors.  If \(K\) is fixed, the limit is the spatial median of finitely many Gaussian vectors and is generally non-Gaussian.  If \(K\to\infty\), the spatial median has a Gaussian limit, but the finite-block median \(s_{\alpha,b}\) can produce a centering term \(\sqrt K\,s_{\alpha,b}\).  The simulation is designed to make this distinction visible.

We compare three regimes.  In the fixed-\(K\) regime, \(K=5\) is held fixed while the block size \(b\) increases.  In the growing-\(K\) regime, both \(K\) and \(b\) increase.  In the aggressive regime, \(K=200\) and \(b=200\), so the number of nodes is large relative to the block size.  For each setting, we record the first coordinate of the root-\(n\)-scaled error and the norm of the root-\(n\)-scaled tangent error.

\begin{figure}[ht!]
    \centering
    \includegraphics[width=\textwidth]{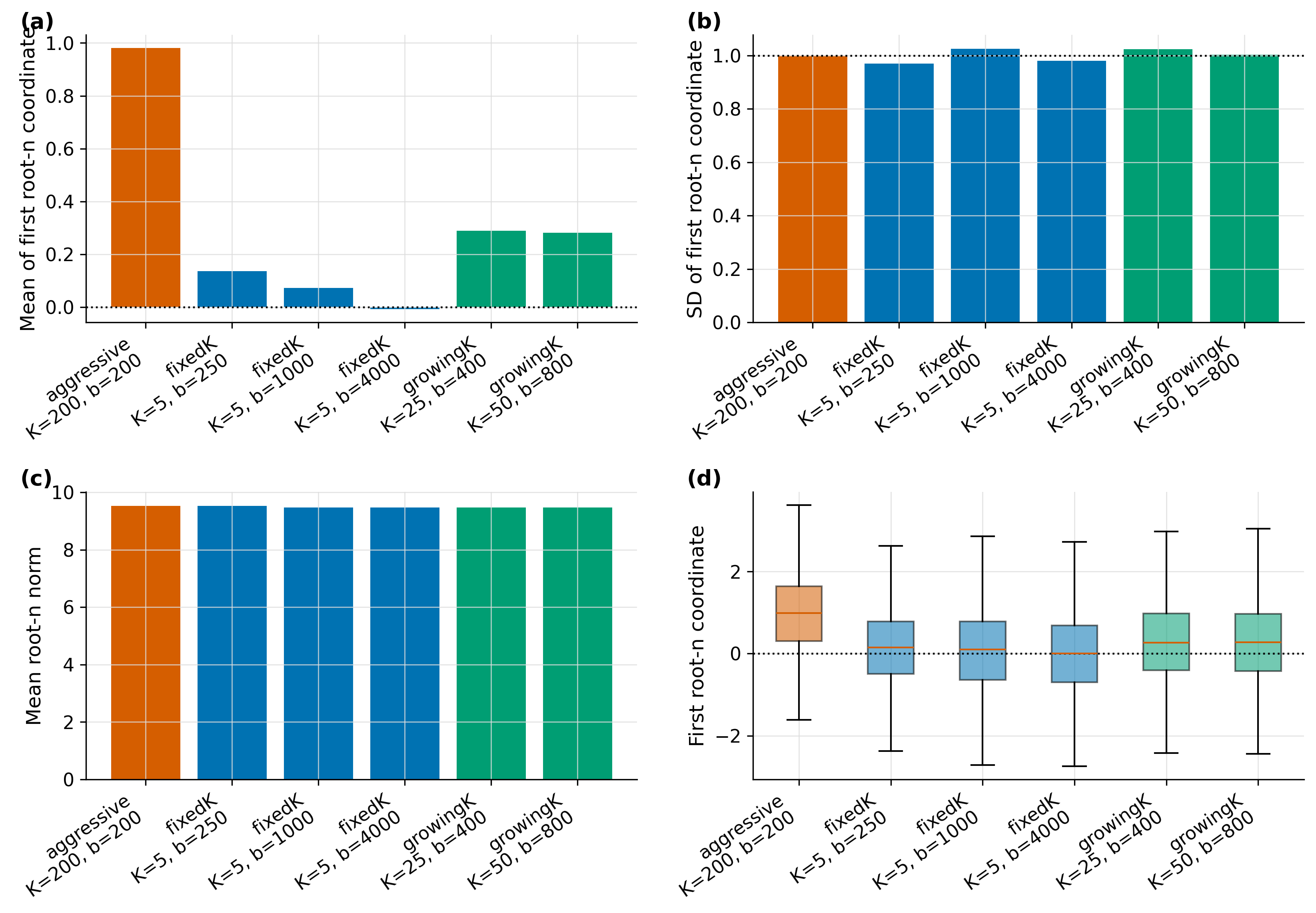}
    \caption{
    {Fixed-\(K\), growing-\(K\), and aggressive-\(K\) regimes.}
    Simulation illustrating the finite-block bias term in the growing-node CLT.  The fixed-\(K\) regime moves toward centered behavior as \(b\) increases, while the aggressive regime has a visibly shifted first root-\(n\) coordinate.  Panels show (a) mean of the first root-\(n\) coordinate, (b) standard deviation of the first root-\(n\) coordinate, (c) mean root-\(n\) norm, and (d) boxplots of the first root-\(n\) coordinate.
    }
    \label{fig:supp-s2-regimes}
\end{figure}

\begin{table}[ht!]
\centering
\caption{
{Asymptotic-regime summary.}
Summary of the first coordinate and norm of the root-\(n\)-scaled tangent error under fixed-\(K\), growing-\(K\), and aggressive-\(K\) regimes.  The aggressive regime has a large nonzero first-coordinate mean, illustrating the finite-block bias effect when the number of nodes is too large relative to the block size.
}
\label{tab:supp-s2-summary}
\begin{tabular}{lrrrrr}
\toprule
Regime & \(K\) & \(b\) & Mean first coord. & SD first coord. & Mean norm \\
\midrule
Aggressive & 200 & 200 & 0.9809 & 0.9990 & 9.5282 \\
Fixed \(K\) & 5 & 250 & 0.1363 & 0.9691 & 9.5304 \\
Fixed \(K\) & 5 & 1000 & 0.0721 & 1.0254 & 9.4642 \\
Fixed \(K\) & 5 & 4000 & -0.0087 & 0.9806 & 9.4641 \\
Growing \(K\) & 25 & 400 & 0.2896 & 1.0237 & 9.4718 \\
Growing \(K\) & 50 & 800 & 0.2812 & 1.0030 & 9.4757 \\
\bottomrule
\end{tabular}
\end{table}

Figure~\ref{fig:supp-s2-regimes} and Table~\ref{tab:supp-s2-summary} show the main patterns.  In the fixed-\(K\) regime, increasing \(b\) moves the mean of the first root-\(n\) coordinate toward zero: it decreases from \(0.136\) at \(b=250\) to \(0.072\) at \(b=1000\), and to approximately \(-0.009\) at \(b=4000\).  This is consistent with the finite-block bias becoming negligible as the node size grows.  In the growing-\(K\) settings, the first-coordinate mean remains around \(0.28\), indicating a small but visible centering effect.  The aggressive setting has a much larger first-coordinate mean, approximately \(0.981\), showing that centered root-\(n\) inference can fail when \(K\) grows too aggressively relative to \(b\).

The standard deviation of the first root-\(n\) coordinate remains close to one across regimes, as shown in panel (b).  The root-\(n\) norm is also similar across settings, with means around \(9.45\)--\(9.53\).  Thus the main visible difference is not a change in overall spread, but a change in centering.  This supports the bias decomposition in Theorem~4: the aggregation noise can have the expected order while the finite-block median contributes a nonzero centering term.

\begin{table}[ht]
\centering
\caption{
Quantiles of root-\(n\)-scaled errors.
Quantiles of the first coordinate and norm of the root-\(n\)-scaled tangent error.  The aggressive regime has a shifted median and upper quantile for the first coordinate, while the norm quantiles remain comparable across regimes.
}
\label{tab:supp-s2-quantiles}
\begin{tabular}{lrrrrrrr}
\toprule
Regime & \(K\) & \(b\) & \(q_{.05}\) first & \(q_{.50}\) first & \(q_{.95}\) first & \(q_{.50}\) norm & \(q_{.95}\) norm \\
\midrule
Aggressive & 200 & 200 & -0.6840 & 0.9895 & 2.6286 & 9.5166 & 10.7496 \\
Fixed \(K\) & 5 & 250 & -1.4655 & 0.1479 & 1.7237 & 9.5137 & 10.6936 \\
Fixed \(K\) & 5 & 1000 & -1.6595 & 0.0974 & 1.7033 & 9.4588 & 10.6199 \\
Fixed \(K\) & 5 & 4000 & -1.6448 & 0.0036 & 1.5551 & 9.4530 & 10.6230 \\
Growing \(K\) & 25 & 400 & -1.3481 & 0.2663 & 1.9886 & 9.4403 & 10.7253 \\
Growing \(K\) & 50 & 800 & -1.3692 & 0.2778 & 1.9416 & 9.4647 & 10.6184 \\
\bottomrule
\end{tabular}
\end{table}

\subsection{Summary}

The supplementary simulations reinforce the two most technical parts of the paper.  The oracle Gaussian node-error experiment confirms that the explicit covariance formula \(V_\alpha\) accurately predicts the empirical covariance of the scaled spatial median.  The relative trace discrepancy is small across variance ratios, correlations, and scales.  The asymptotic-regime experiment shows that finite-block bias is not merely a proof artifact: when the number of nodes grows too aggressively relative to the node size, the root-\(n\)-scaled estimator can have a nonzero centering shift.  Together, these experiments support the covariance and bias decompositions used in the main theoretical development.

\section{Additional real-data analyses}
\label{sec:supp-real-data}

This section reports two additional real-data analyses.  The main paper focuses on the 10x mouse brain data because it is a large single-cell RNA-seq example where PCA is a standard preprocessing and visualization tool.  Here we use two supplementary datasets to examine whether the same mean--subspace aggregation behavior appears in different settings.  The first is a human single-cell dataset from Tabula Sapiens \citep{thetabulasapiensconsortium_2022_TabulaSapiensMultipleorgan}, which stays close to the single-cell motivation but introduces a different biological context.  The second is a climate EOF/PCA example based on NOAA OISST sea-surface temperature fields \citep{reynolds_2002_ImprovedSituSatellite, huang_2021_ImprovementsDailyOptimum}, which tests the method outside genomics.

The purpose of these analyses is not to claim that one method is uniformly best in all real-data settings.  Rather, the goal is to examine whether the empirical patterns from the main paper persist: random-subset PCA is unstable, distributed aggregation improves stability, fixed-scale MoM and scale-calibrated MoM produce competitive robust summaries, and the selected scale reflects node-level mean--subspace uncertainty.  As in the main real-data example, full-data PCA is treated as a centralized reference, not as ground truth.

\subsection{Tabula Sapiens blood cells}
\label{subsec:supp-tabula}

We first analyzed the blood component of the Tabula Sapiens human single-cell atlas.  Tabula Sapiens is a multi-organ single-cell reference atlas with rich biological metadata, and the blood subset provides a useful supplementary example because it is large enough to require distributed processing while remaining computationally manageable.  We selected \(p=2000\) highly variable genes from the log-normalized expression matrix and compared distributed PCA summaries to the full-data PCA reference.  The analysis used \(n=50115\) cells, \(K\in\{20,50\}\) nodes, and ranks \(r\in\{10,20\}\).  For each node, we computed the node mean and node principal subspace, then compared projector averaging, random subset PCA, fixed-scale MoM PCA with \(\alpha=1\), and scale-calibrated MoM PCA.  Results were averaged over 10 random shardings.

\begin{figure}[ht!]
    \centering
    \includegraphics[width=\textwidth]{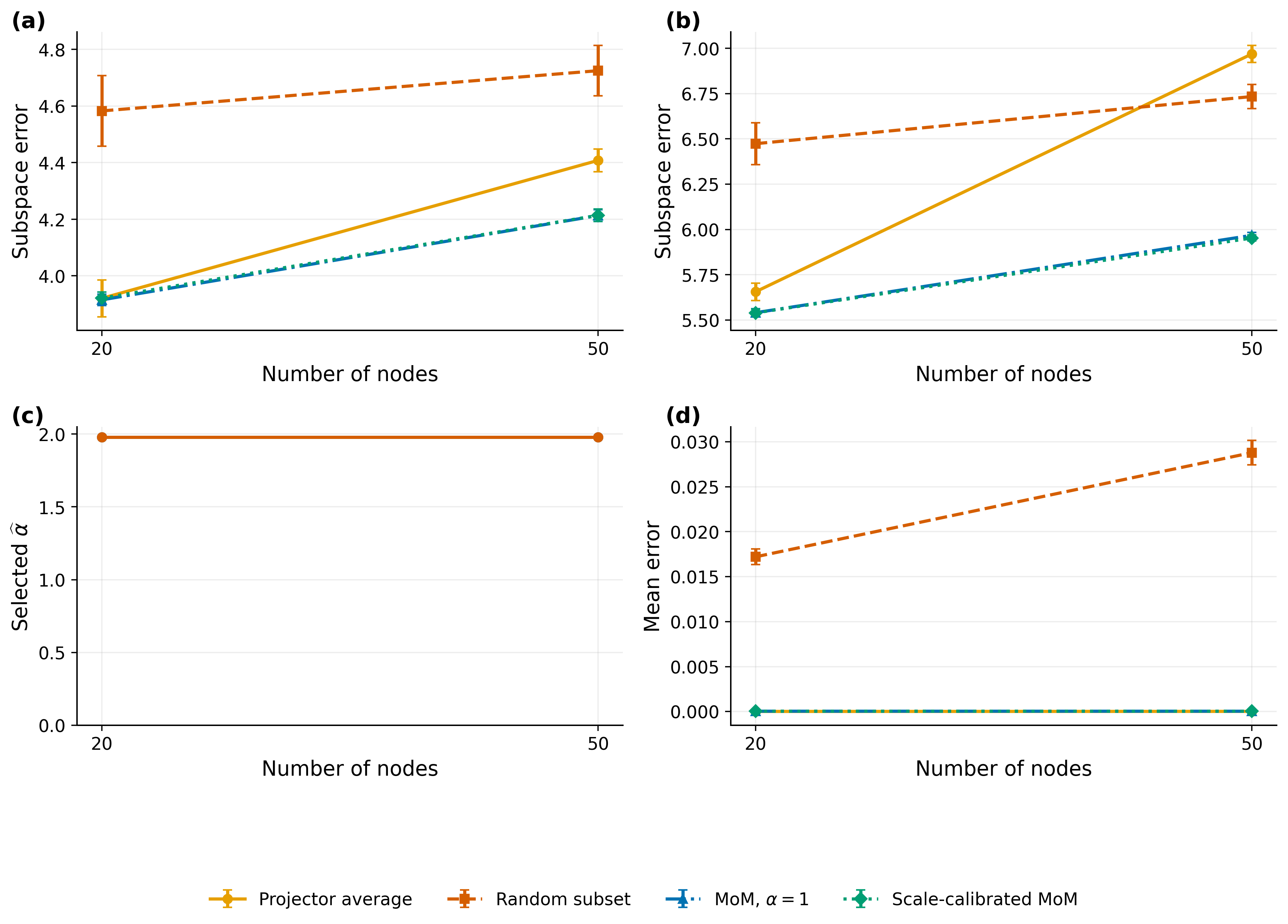}
    \caption{
    {Tabula Sapiens blood-cell PCA.}
    Distributed PCA comparison on the Tabula Sapiens blood subset using \(p=2000\) highly variable genes, \(K\in\{20,50\}\) nodes, and ranks \(r\in\{10,20\}\).  Full-data PCA is used as the centralized reference.  Random subset PCA uses one node-sized subsample, while projector averaging and the two MoM procedures aggregate all node estimates.  Panels show (a) subspace error for \(r=10\), (b) subspace error for \(r=20\), (c) selected calibrated scale, and (d) mean error.
    }
    \label{fig:supp-tabula}
\end{figure}

Figure~\ref{fig:supp-tabula} summarizes the results.  Random subset PCA has the largest mean error and is consistently unstable in the subspace coordinate.  This is expected because it discards most nodes and uses only one node-sized subsample.  Projector averaging is highly accurate for the mean, but its subspace error increases with the number of nodes.  The MoM estimators have slightly larger mean errors than projector averaging, but their mean errors remain very small on the scale of the log-normalized data.  In contrast, their subspace errors are competitive and often smaller than projector averaging, especially for \(r=20\).

\begin{table}[ht!]
\centering
\caption{
{Tabula Sapiens blood-cell analysis.}
Representative results for \(K=50\) nodes using \(p=2000\) highly variable genes.  Errors are measured relative to the full-data PCA reference.  Random subset PCA uses one node-sized subsample, projector averaging is the standard distributed baseline, and the MoM estimators aggregate node mean--subspace pairs on \(\R^p\times\Gr(r,p)\).
}
\label{tab:supp-tabula}
\begin{tabular}{lcccc}
\toprule
Rank & Method & Mean error & Subspace error & \(\widehat\alpha\) \\
\midrule
\multirow{4}{*}{\(r=10\)}
& Projector average & \(0.000018\) & \(4.4080\) & -- \\
& Random subset & \(0.028788\) & \(4.7250\) & -- \\
& MoM, \(\alpha=1\) & \(0.000036\) & \(4.2133\) & \(1.000\) \\
& Scale-calibrated MoM & \(0.000035\) & \(4.2141\) & \(1.980\) \\
\midrule
\multirow{4}{*}{\(r=20\)}
& Projector average & \(0.000025\) & \(6.9695\) & -- \\
& Random subset & \(0.027535\) & \(6.7340\) & -- \\
& MoM, \(\alpha=1\) & \(0.000042\) & \(5.9675\) & \(1.000\) \\
& Scale-calibrated MoM & \(0.000042\) & \(5.9525\) & \(1.980\) \\
\bottomrule
\end{tabular}
\end{table}

Table~\ref{tab:supp-tabula} reports representative values for \(K=50\).  For \(r=10\), fixed-scale MoM and scale-calibrated MoM both reduce the subspace error relative to projector averaging and random subset PCA.  For \(r=20\), the gap is more pronounced: projector averaging has subspace error about \(6.97\), while fixed-scale MoM and scale-calibrated MoM have subspace errors about \(5.97\) and \(5.95\), respectively.  The calibrated scale is close to the upper endpoint, \(\widehat\alpha_{\rm rPCA}\approx1.98\), for both ranks.  This indicates large node-level Grassmann dispersion relative to the Euclidean mean dispersion, causing the calibration rule to downweight the subspace component in the aggregation metric.

These results support the interpretation from the main paper.  Scale calibration is primarily an inferential and geometric balancing mechanism.  In this Tabula Sapiens analysis, the selected scale suggests that node subspaces are substantially more variable than node means.  The calibrated estimator remains close to the fixed-scale MoM point estimate, while its selected scale records the node-level mean--subspace uncertainty imbalance.

\subsection{NOAA OISST sea-surface temperature EOF analysis}
\label{subsec:supp-oisst}

We next considered a climate example based on NOAA Optimum Interpolation Sea Surface Temperature (OISST).  Empirical orthogonal function analysis is the climate-science analogue of PCA, so this example provides a non-genomic check of the proposed aggregation framework.  The OISST product is a gridded sea-surface temperature record constructed from satellite and in-situ observations.  We formed an anomaly matrix from a regional, coarsened OISST field and treated time points as observations.  The analysis used \(n=120\) time points, \(p=80\) spatial coordinates, \(K=5\) temporal nodes, and rank \(r=3\).  Because the data are temporally dependent, this analysis should be interpreted as a cross-domain robustness and scalability illustration rather than a direct validation of the independent-sampling theory.

\begin{figure}[ht!]
    \centering
    \includegraphics[width=\textwidth]{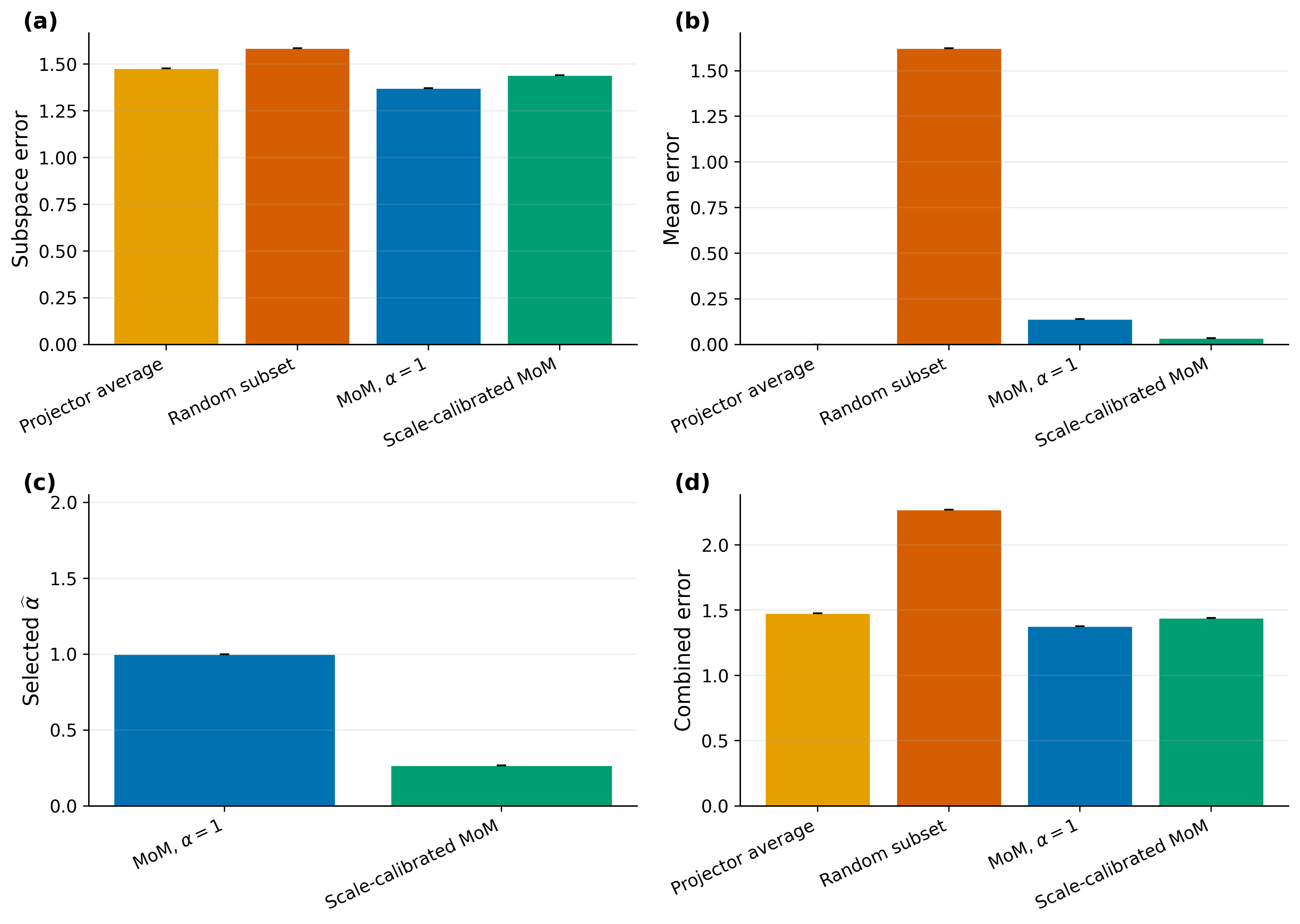}
    \caption{
    {NOAA OISST EOF/PCA analysis.}
    Distributed EOF/PCA comparison on a regional OISST sea-surface-temperature anomaly matrix.  Full-data EOF/PCA is used as the centralized reference.  Projector averaging, random subset PCA, fixed-scale MoM, and scale-calibrated MoM are compared using \(K=5\) temporal nodes and rank \(r=3\).  Panels show (a) subspace error, (b) mean error, (c) selected calibrated scale, and (d) combined error.
    }
    \label{fig:supp-oisst}
\end{figure}

\begin{table}[ht!]
\centering
\caption{
{NOAA OISST EOF/PCA analysis.}
Results for the regional OISST anomaly matrix with \(n=120\), \(p=80\), \(K=5\) temporal nodes, and rank \(r=3\).  Errors are measured relative to the full-data EOF/PCA reference.  This example is temporally dependent and is used as a supplementary cross-domain demonstration.
}
\label{tab:supp-oisst}
\begin{tabular}{lccc}
\toprule
Method & Mean error & Subspace error & \(\widehat\alpha\) \\
\midrule
Projector average & \(0.0000\) & \(1.4769\) & -- \\
Random subset & \(1.6245\) & \(1.5865\) & -- \\
MoM, \(\alpha=1\) & \(0.1383\) & \(1.3713\) & \(1.000\) \\
Scale-calibrated MoM & \(0.0352\) & \(1.4411\) & \(0.268\) \\
\bottomrule
\end{tabular}
\end{table}

Figure~\ref{fig:supp-oisst} and Table~\ref{tab:supp-oisst} summarize the comparison.  Projector averaging has essentially zero mean error because the node means average back to the full-data mean in this partitioned setting.  Its subspace error, however, is \(1.48\).  Fixed-scale MoM with \(\alpha=1\) gives the smallest subspace error, about \(1.37\), while scale-calibrated MoM gives subspace error \(1.44\).  The random subset baseline is worst overall, with mean error \(1.62\) and subspace error \(1.59\).

The calibrated scale is \(\widehat\alpha_{\rm rPCA}\approx0.27\), indicating that the node-level mean dispersion dominates the per-dimension calibration.  This choice improves the mean error relative to fixed-scale MoM, reducing it from \(0.138\) to \(0.035\), but does not minimize subspace error.  This is consistent with the main paper's message: calibration is a data-adaptive mean--subspace balancing rule rather than a procedure that uniformly minimizes every component loss.  In the OISST example, fixed-scale MoM is preferable for recovering the EOF subspace, while scale-calibrated MoM gives a more mean-stable robust aggregate.

\subsection{Summary}

The supplementary real-data analyses reinforce two practical lessons.  First, random subset PCA is an unreliable baseline when the goal is to approximate a full-data mean--subspace summary; it often has much larger mean error or subspace error because it ignores most nodes.  Second, robust MoM aggregation can improve subspace stability relative to simple distributed baselines, but the scale-calibrated version should be interpreted through the mean--subspace tradeoff.  In Tabula Sapiens, the calibration identifies large node-level subspace dispersion and selects \(\widehat\alpha_{\rm rPCA}\) close to \(2\).  In OISST, the calibration selects a smaller value of \(\alpha\), improving mean stability while giving a modestly larger subspace error than fixed-scale MoM.  Together with the main 10x mouse brain analysis, these examples show that scale calibration adapts to the empirical node-error geometry rather than enforcing a fixed universal preference for either the mean or the subspace component.

\end{document}